\documentclass[a4paper,11pt]{article} 
\pdfoutput=1
\usepackage{jheppub} 
\usepackage{multirow}

\usepackage{bbm}
\usepackage{longtable,booktabs,graphicx}

\allowdisplaybreaks

\title{\boldmath Covariant orbital-spin scheme for any spin based on irreducible tensor}

\author[a]{Hao-Jie Jing,}
\author[b,c]{Di Ben,}
\author[a,c]{Shu-Ming Wu,}
\author[a]{Jia-Jun Wu,}
\author[a,c,d]{and Bing-Song Zou}

\affiliation[a]{School of Physical Sciences, University of Chinese Academy of Sciences (UCAS), \\Beijing 100049, China}
\affiliation[b]{School of Nuclear Science and Technology,
University of Chinese Academy of Sciences (UCAS), \\ Beijing 101408, China}
\affiliation[c]{CAS Key Laboratory of Theoretical Physics, Institute of Theoretical Physics, \\Chinese Academy of Sciences, \\ Zhong Guan Cun East Street 55, Beijing 100190, China}
\affiliation[d]{Southern Center for Nuclear-Science Theory (SCNT),
Institute of Modern Physics, \\Chinese Academy of Sciences, \\Huizhou 516000, China}

\emailAdd{jinghaojie@ucas.ac.cn}
\emailAdd{bendi20@mails.ucas.ac.cn}
\emailAdd{wushuming@itp.ac.cn}
\emailAdd{wujiajun@ucas.ac.cn}
\emailAdd{zoubs@itp.ac.cn}

\abstract{
In hadron spectrum physics, the partial wave analysis is a primary method used to extract properties of hadronic resonances. 
The covariant orbital-spin coupling scheme holds unique advantages over other partial wave methods due to its Lorentz covariant form and determined orbital-spin quantum numbers.
This paper presents a general form of the covariant orbital-spin coupling scheme based on the irreducible tensor of the homogeneous proper Lorentz group and its little groups.
A systematic procedure for constructing partial wave amplitude in a Lorentz covariant way is provided, which can be applied to both massive and massless particles.
Specific examples are also included.
}

\begin{document} 
\maketitle
\flushbottom

\section{Introduction}
\label{sec:intro}
Quantum chromodynamics (QCD) is a fundamental theory to describe strong interaction between quarks and gluons. 
However, due to its non-perturbative properties under the hadron energy scale, one cannot get the hadron-hadron interaction through QCD in the framework of perturbation theory,
i.e., it is unavailable to extract the hadron-hadron interaction directly from quarks and gluons. 
Until now, since the analytical mathematical method for non-perturbation is still missing, it is sufficient for the effective way to extract the hadron information from experimental data.
In phenomenology, establishing a clear hadron spectrum is the promise to understand the low-energy behavior of strong interaction, including the properties of hadron mass, lifetime or width, spin, parity and so on. 
In order to obtain such informations, it is necessary to analyze the final state invariant mass and angular distributions by partial wave analysis (PWA) which is, in principle, a model-independent approach.
PWA is a standard method that extracts quantum numbers from invariant mass and angular distributions. It projects the scattering amplitude into several parts with definite orbital angular momentum $L$ and spin $S$ quantum numbers.

There are several PWA formalisms, including non-covariant methods such as the helicity formalism~\cite{Jacob:1959at}, the Zemach formalism~\cite{Zemach:1965ycj}, multipole analysis~\cite{Wennstrom:1968mkp,Cutkosky:1979zv,Arndt:1989ww,Hanstein:1997tp}, and covariant methods such as the covariant helicity formalism~\cite{PhysRevD.48.1225}, the covariant effective Lagrangian approach~\cite{Liang:2002tk}, and the covariant orbital-spin ($L$-$S$) scheme~\cite{Zou:2002ar,Zou:2002yy,Dulat:2005in,Dulat:2011rn}.
Compared to non-covariant methods, covariant methods maintain Lorentz covariance, making it easier to apply these schemes in various cascade decays.
Additionally, fixed $L$-$S$ quantum numbers of the covariant $L$-$S$ scheme help distinguish contributions from different partial wave amplitudes and make it simpler to introduce $L$-dependent form factors.
The covariant $L$-$S$ scheme was first proposed in refs.~\cite{Zou:2002ar,Zou:2002yy} to describe $\psi MM$, $N^*NM$ partial wave amplitudes and later applied to radiation decay processes such as $N^*N \gamma$ and $\psi M \gamma$~\cite {Dulat:2005in,Dulat:2011rn}, which is widely used by BESIII group for PWA~\cite{BESIII:2022chl,Ge:2022dzi,Lu:2022zjl,BESIII:2022vaf,BESIII:2022npc}.

However, in previous works~\cite{Zou:2002ar,Zou:2002yy,Dulat:2005in,Dulat:2011rn} of covariant $L$-$S$ scheme, a general formula for partial wave amplitude with any spin $s=N/2~(N=0,1,2,\cdots)$ is lacking.
Especially for the high half-integer spin cases, e.g., the coupling of one meson and two fermions with spin-$\frac32$ is missing.
In this paper, a general theoretical framework of covariant $L$-$S$ scheme has been developed by using irreducible tensors (IRTs) of the homogeneous proper Lorentz group ($L_p$) and its little groups.

This paper is organized as follows.
In section~\ref{sec:covariant_LS_irten}, we systematically introduce a general form of three-particle Lorentz covariant partial wave amplitude.
In section~\ref{sec:app}, we provide three examples, including (a) a process with three bosons; (b) a process involving two fermions; and (c) a process that includes a massless particle.
In section~\ref{sec:summary}, a brief summary is given.

\section{The general framework of covariant orbital-spin scheme}
\label{sec:covariant_LS_irten}


\subsection{General form of covariant $L$-$S$ scheme}
\label{sec:covariant_General}

In this section, we will make a general discussion on Lorentz covariant partial wave amplitudes. 
To ensure the covariance of scattering amplitudes under Lorentz transformations, the coupling structures (amplitudes without spin wave functions of external particles) of $n$-particle vertex amplitudes must be an order-$n$ covariant tensor (COVT).
Any COVT can be decomposed into IRTs, and any IRT can be expressed by order-3 IRTs. For a brief discussion on this topic, please refer to appendix~\ref{appd:irrep_and_irten}.
From the physics point of view, the multi-particle vertex can be derived recursively from the three-particle vertex. 
Therefore, in this work, we will focus on the coupling structure involving three external particles.

The vertex amplitude of three-particle interaction (e.g., consider a process of $1\to 2+3$)\footnote{It is important to note that a Lorentz covariant three-particle vertex amplitude can represent various processes, including $1\to2+3$ and $2\to1+3$, due to the crossing symmetry. However, when examining the partial wave amplitude with a specific Lorentz covariant coupling structure, the partial wave components in the amplitudes of different processes connected by crossing symmetry are typically distinct. For some examples, please refer to appendix~\ref{appd_LS_decomposition}.} 
with arbitrary spin can be written in the following general form,
\begin{equation}
	{\cal A}_{\sigma_1}^{\sigma_2\sigma_3}\left(\mathbf{p}_1,\mathbf{p}_2,\mathbf{p}_3;s_1,s_2,s_3\right) = \Gamma^{\alpha_2\alpha_3}_{\alpha_1}\left(\mathbf{p}_1,\mathbf{p}_2,\mathbf{p}_3\right) \,\bar{u}_{\sigma_1}^{\alpha_1}(\mathbf{p}_1;s_1)\, u^{\sigma_2}_{\alpha_2}(\mathbf{p}_2;s_2)\, u^{\sigma_3}_{\alpha_3}(\mathbf{p}_3;s_3),
	\label{eq:any_spin_j_general}
\end{equation}
where $u^{\sigma_i}_{\alpha_i}(\mathbf{p}_i;s_i)/\bar{u}_{\sigma_i}^{\alpha_i}(\mathbf{p}_i;s_i)$ is the covariant/contravariant spin wave function of the $i$-th particle with total spin $s_i$, spin polarization component $\sigma_i$, three momentum $\mathbf{p}_i$
and Lorentz covariant index $\alpha_i$  carrying the representation $[\alpha_i]$ of the homogeneous proper Lorentz group ($L_p$);
$\Gamma^{\alpha_2\alpha_3}_{\alpha_1}(\mathbf{p}_1,\mathbf{p}_2,\mathbf{p}_3)$ is an order-3 COVT; we adopt Einstein's summation rule, the repeated indices indicate summation. 
For convenience, we will drop $s_1$, $s_2$ and $s_3$ in ${\cal A}$ in the rest of this paper.

The $L$-$S$ scheme aims to decompose the full amplitude into terms with fixed orbital angular momentum ($L$) and total spin ($S$) of two final-state particles. This can be achieved in three steps.
At the first step, we need to separate the continuous part labeled by momentum $\mathbf{p}$ from the discrete part labeled by spin $s$ in the spin wave function as follows,
\begin{align}
	u^{\sigma}_{\alpha}(\mathbf{p};s)=     D_{\alpha}^{~~\beta}\left(h_{\mathbf{p}}\right)\,
	u^{\sigma}_{\beta}(\mathbf{k};s),
	\label{eq:any_spin_general_U}
\end{align}
where $D_{\alpha}^{~~\beta}\left(g\right)\,(g\in L_p)$ is a Lorentz transformation matrix; $h_\mathbf{p}\left(\in L_p\right)$ is a pure-boost transformation; 
$u^{\sigma}_{\alpha}(\mathbf{k};s)$ is the spin wave function in the standard momentum frame\footnote{For further details about standard momentum frame, please refer to section V of chapter II in ref.~\cite{weinberg_1995}.}. 
The standard momentum is labeled as $k^{\mu}=\left(k^0,\mathbf{k}\right)$.
As a result, $u^{\sigma}_{\alpha}(\mathbf{k};s)$ does not contain any continuous degrees of freedom, which just describes a particle's intrinsic property, spin.
Since $S$ and $L$ are defined in the standard momentum frame of particle-$1$, i.e., $\mathbf{p}_1=\mathbf{k}_1$, the eq.~\eqref{eq:any_spin_j_general} becomes
\begin{align}
	{\cal A}_{\sigma_1}^{\sigma_2\sigma_3}(\mathbf{k}_1,\mathbf{p}^*_2,\mathbf{p}^*_3) =&\, \underbrace{\Gamma^{\alpha_2\alpha_3}_{\alpha_1}(\mathbf{k}_1,\mathbf{p}^*_2,\mathbf{p}^*_3) \,D_{\alpha_2}^{~~\beta_2}\left(h_{\mathbf{p}^*_2}\right)\, D_{\alpha_3}^{~~\beta_3}\,\left(h_{\mathbf{p}^*_3}\right)}_{\text{pure-orbitial part}}~\times\notag\\
	&\,\underbrace{\bar{u}_{\sigma_1}^{\alpha_1}(\mathbf{k}_1;s_1)\,u^{\sigma_2}_{\beta_2}(\mathbf{k}_2;s_2)\,u^{\sigma_3}_{\beta_3}(\mathbf{k}_3;s_3)}_{\text{pure-spin part}},
	\label{eq:any_spin_general_LS_1_rest}
\end{align} 
where $\mathbf{p}^*_2$ and $\mathbf{p}^*_3$ are the three-momenta of particle-2 and particle-3 in the standard momentum frame of particle-1, respectively. 

At the second step, we will decompose the above amplitude (eq.\eqref{eq:any_spin_general_LS_1_rest}) into the pure orbital angular momentum $L$ part and pure spin angular momentum $S$ part.
The pure-spin part can be kept but we need spin projection tensors to pick out fixed $S$ of two final-state particles, while the pure-orbital part will be divided by fixed $L$.
Consequently, a Lorentz covariant partial wave amplitude can be expressed as follows,
\begin{align}
	{\cal A}_{\sigma_1}^{\sigma_2\sigma_3}\left(\mathbf{k}_1,\mathbf{p}^*_2,\mathbf{p}^*_3;L,S\right)=\, \Gamma_{\alpha_1}^{\alpha_2\alpha_3}(\mathbf{k}_1,\mathbf{p}^*_2,\mathbf{p}^*_3;L,S)\,
	\bar{u}_{\sigma_1}^{\alpha_1}(\mathbf{k}_1;s_1)\,u^{\sigma_2}_{\alpha_2}(\mathbf{k}_2;s_2)\,u^{\sigma_3}_{\alpha_3}(\mathbf{k}_3;s_3),
	\label{eq:general_LS_expand_1_rest}
\end{align}
where 
$\Gamma_{\alpha_1}^{\alpha_2\alpha_3}(\mathbf{k}_1,\mathbf{p}^*_2,\mathbf{p}^*_3;L,S)$ is a Lorentz covariant coupling structure with definite $L$ and $S$ quantum numbers.
The main objective of this paper is to express the coupling structure explicitly.
In ref.~\cite{Zou:2002yy}, a pure spin wave function for two fermions was constructed by combining various Lorentz structures.
For instance, one can construct a pure spin-1 system $\left({}^3S_1\right)$ of two fermions with spin-$\frac{1}{2}$ by a linear combination of $\bar{\psi}_2\gamma_{\mu}\psi_1$ and $\bar{\psi}_2\stackrel{\leftrightarrow}{\partial}_{\mu}\psi_1$ by ignoring the ${}^3P_0$ component (see table~\ref{tab:12_12_j_common_results}).
Therefore, it is feasible to use IRTs to decompose the amplitudes containing different Lorentz structures, and obtain partial wave amplitudes through their linear combinations. 
However, with the increase of $L$ and $S$, the amount of computation will increase dramatically, see table~\ref{tab:12_32_j_common_results} for the case including a fermion with spin-$\frac{3}{2}$. 
For this reason, here we directly consider how to construct the Lorentz covariant partial wave amplitude based on the IRTs of $L_p$ and the IRTs of the little group SO(3).
At the third step, since three-particle amplitude is just a block for the whole amplitude in the cascade reaction, it is necessary to transfer the rest frame of initial state to any given frame, i.e., from 
$(\mathbf{k}_1, \mathbf{p}^*_2,\mathbf{p}^*_3)$ to $(\mathbf{p}_1, \mathbf{p}_2,\mathbf{p}_3)$.
It is quiet straightforward to use several Lorentz-boost transformations. Finally, one will get a Lorentz covariant partial wave amplitude ${\cal A}_{\sigma_1}^{\sigma_2\sigma_3}\left(\mathbf{p}_1,\mathbf{p}_2,\mathbf{p}_3;L,S\right)$ in any frame as follows,
\begin{align}
	{\cal A}_{\sigma_1}^{\sigma_2\sigma_3}(\mathbf{p}_1,\mathbf{p}_2,\mathbf{p}_3;L,S) =&\,\Gamma_{\alpha_1}^{\alpha_2\alpha_3}(\mathbf{k}_1,\mathbf{p}^*_2,\mathbf{p}^*_3;L,S)\,
	D_{\alpha_2}^{~~\beta_2}\left(R_{12}\right)
	D_{\alpha_3}^{~~\beta_3}\left(R_{13}\right)\notag\\
	&\,\times\bar{u}_{\sigma_1}^{\alpha_1}(\mathbf{k}_1;s_1)~u^{\sigma_2}_{\beta_2}(\mathbf{k}_2;s_2)~u^{\sigma_3}_{\beta_3}(\mathbf{k}_3;s_3),
	\label{eq:general_LS_expand_1_motion}
\end{align}
where $R_{1i}=h_{\mathbf{p}_i^*}^{-1} \cdot h_{\mathbf{p}_1}^{-1} \cdot h_{\mathbf{p}_i}~(i=2,3)$.

To construct Lorentz covariant partial wave amplitudes based on eq.~\eqref{eq:general_LS_expand_1_motion}, one needs to write down the spin wave functions of particles with arbitrary spin, and the specific form of covariant coupling structure $\Gamma_{\alpha_1}^{\alpha_2\alpha_3}(\mathbf{k}_1,\mathbf{p}^*_2,\mathbf{p}^*_3;L,S)$ which can be organized as three parts, (a) the covariant tensor of orbital angular momentum $L$ part, (b) the covariant tensor of $s_2$ and $s_3$ coupled to total spin $S$, and (c) the covariant tensor of $S$ and $L$ coupled to $s_1$, as follows,
\begin{align} 	\Gamma_{\alpha_1}^{\alpha_2\alpha_3}(\mathbf{k}_1,\mathbf{p}^*_2,\mathbf{p}^*_3;L,S) =	P^{\alpha^L\alpha^S}_{\alpha_1}(\mathbf{k}_1;s_1,L,S)~P^{\alpha_2\alpha_3}_{\alpha^S}(\mathbf{k}_1;S,s_2,s_3)~\tilde{t}^{(L)}_{\alpha^L}(\mathbf{k}_1,\mathbf{p}^*_2-\mathbf{p}^*_3),	\label{eq:coupling_structure_Gamma_1_rest}
\end{align} 
where $P^{\alpha^L\alpha^S}_{\alpha^J}(\mathbf{k}_1;J,L,S)$ is the angular momentum coupling structure for $\mathbf{J}=\mathbf{L}+\mathbf{S}$ in the standard momentum frame of particle-1; 
and $\tilde{t}^{(L)}_{\alpha^L}(\mathbf{k}_1,\mathbf{p}^*_2-\mathbf{p}^*_3)$ is the $L$-wave orbital angular tensor.
Typically, these tensors are the IRTs of the group SO(3) with Lorentz covariance, which can be constructed by spin wave functions according to the eigen-function method~\cite{Chen:1985zzd}.
Thus, all building blocks of covariant $L$-$S$ scheme can be reduced to the spin wave functions, while the spin wave function can be described by the representation of $L_p$.

The remaining part of this section is organized as: in subsection~\ref{sec:any_spin_wave_function}, we give a general discussion on Lorentz covariant spin wave function for any spin;
in subsection~\ref{sec:irten_Lp}, we show how to use these spin wave functions to construct the IRTs of $L_p$ and the little group SO(3); 
in subsection~\ref{sec:massless_spin_wave_func}, we discuss the difference between spin wave functions for massive and massless particles;
in subsection~\ref{sec:irten_exact}, we explicitly present a general expression of Lorentz covariant partial wave amplitude ${\cal A}_{\sigma_1}^{\sigma_2\sigma_3}\left(\mathbf{p}_1,\mathbf{p}_2,\mathbf{p}_3;L,S\right)$.
%


\subsection{Spin wave function based on irreducible representations of $L_p$ }
\label{sec:any_spin_wave_function}

By considering Poinc$\acute{\text{a}}$re invariance and causality, Weinberg had constructed the covariant causal field operator for both massive and massless particles with spin $s$~\cite{Weinberg:1964cn,Weinberg:1964ev,Weinberg:1969di}, the covariant and contravariant spin wave functions with momentum $p^\mu=(\omega_{\mathbf{p}},\mathbf{p})~\left(\omega_{\mathbf{p}}=\sqrt{|\mathbf{p}|^2+m^2}\right)$ are defined as follows,
\begin{equation}
	u^\sigma_\alpha(\mathbf{p};s)=D_{\alpha}^{~~\beta}(h_\mathbf{p})\,u^\sigma_\beta(\mathbf{k};s),\qquad \bar{u}_\sigma^\alpha(\mathbf{p};s)=D_{\beta}^{~~\alpha}(h_\mathbf{p}^{-1})\,\bar{u}_\sigma^\beta(\mathbf{k};s).
	\label{eq:move_wave_function}
\end{equation}
The Lorentz transformation matrix in any representation $[\alpha]$ can be derived easily, please refer to appendix~\ref{appd:Lp_transformation} for a brief discussion.
The contraction of a covariant spin wave function with the corresponding contravariant spin wave function obeys the following orthonormal relation,
\begin{equation}
	\bar{u}^{\alpha}_{\sigma_1}(\mathbf{p};s)\, u_{\alpha}^{\sigma_2}(\mathbf{p};s)~=~\delta^{~~\sigma_2}_{\sigma_1}.
	\label{eq:orthogonal_normalization}
\end{equation}

\begin{table}
	\centering
	\begin{tabular}{c|c|c}
		\hline
		~~~Standard $k^\mu$~~~&~~~Little group $L_{p,k}$~~~&~~~Generators of $L_{p,k}$~~~ \\
		\hline
		$(\pm |m|,0,0,0)$ & SO(3) & $J_1,~J_2,~J_3$\\
		$(\pm|\mathbf{k}|,0,0,|\mathbf{k}|)$ & ISO(2) & ~$J_3,~(J_2+K_1),~(J_1-K_2)$~\\
		$(0,0,0,|\mathbf{k}|)$ & SO(1,2) & $J_3,~K_1,~K_2$\\
		$(0,0,0,0)$ & SO(1,3) & $J_1,~J_2,~J_3,~K_1,~K_2,~K_3$\\
		\hline
	\end{tabular}
\caption{
	Standard momenta $k^\mu$ and the corresponding little group $L_{p,k}$ for various classes of four-momenta.
	$J_i\,(i=1,2,3)$ and $K_i\,(i=1,2,3)$ are rotation and boost generators of $L_p$, respectively and satisfy the commutation relations: $[J_i,J_j] = i\epsilon_{ijk} J_k$, $[K_i,K_j] = -i \epsilon_{ijk} J_k$ and $[J_i,K_j] = i \epsilon_{ijk} K_k$.
}
	\label{tab:little_group}
\end{table}

Some systematic discussion would be useful to obtain the specific form of $u^{\sigma}_\alpha(\mathbf{k};s)$ and $\bar{u}^{\sigma}_\alpha(\mathbf{k};s)$.
Firstly, one should fix a representation $[\alpha]$. 
On the one hand, 
by recombining the six generators of $L_p$ (table~\ref{tab:little_group}) as,
\begin{equation}
	A_i  = J_i + i\, K_i, \quad B_i = J_i - i\, K_i~(i=1,2,3).
	\label{eq:def_A_B}
\end{equation}
Then, one has,
\begin{equation}
	[A_i,A_j] = i \epsilon_{ijk}A_k,\qquad
	[B_i,B_j] = i\epsilon_{ijk} B_k,\qquad[A_i,B_j]=0.
	\label{eq:leftA_rightB}
\end{equation}
These commutation relations imply that $L_p \backsimeq \text{SU}(2)_L \otimes \text{SU}(2)_R$. 
Thus, an irreducible representation (IRREP) of $L_p$ can be labeled by a binary $(s_L, s_R)$ where $s_L$ and $s_R$ are any positive integers or half integers. For instance, the IRREP $\left(\frac{1}{2},0\right)$ describes a left-handed Dirac spinor; the IRREP $\left(\frac{1}{2},\frac{1}{2}\right)$ describes a Lorentz four-vector and so on.
On the other hand, in order to conveniently describe a process with definite $\mathcal{P}$-parity (space inversion) and $\mathcal{C}$-parity (charge conjugation), it is better to consider an IRREP $(s_L,s_R)$ together with the corresponding complex conjugate representation $(s_R,s_L)$.
Thus, we will focus on self-conjugate representations in table~\ref{tab:irreps_Lip} and use the following convention,
\begin{equation}
	[s_L,s_R]~\equiv~\left\{\begin{array}{cc}
		~(s_L,s_R)~ &\text{for $s_L=s_R$} \\
		\\
		~(s_L,s_R)\oplus(s_R,s_L)~ &\text{for $s_L \neq s_R$}
	\end{array}\right..
	\label{eq:self_conj_rep_label}
\end{equation}.

\begin{figure}[tbph]
	\centering
	\includegraphics[width=0.7\linewidth]{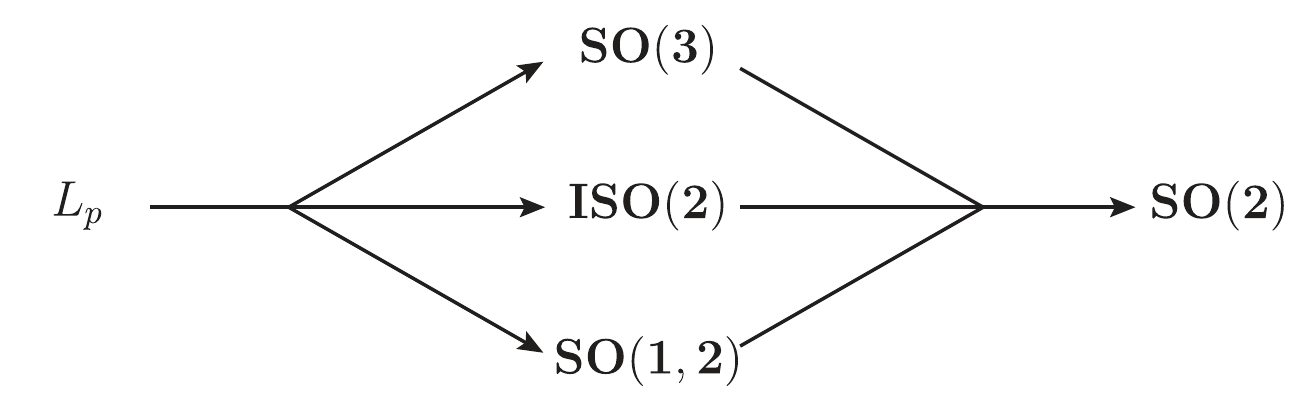}
	\caption{Three different little group chains of the homogeneous Lorentz group $L_p$.}
	\label{fig:properlorentzgroupchain}
\end{figure}

\begin{table}[htbp]
	\centering
	\begin{tabular}{cc}
		\hline
		~~~Representation~~~ \hspace{1cm}&\hspace{1cm} Physical correspondence \\
		\hline
		$\left(0,0\right)$\hspace{1cm}&\hspace{1cm} Scalar \\
		$\left(\frac{1}{2},0\right)\oplus\left(0,\frac{1}{2}\right)$\hspace{1cm}&\hspace{1cm}  Dirac spinor (spin 1/2)\\
		$\left(\frac{1}{2},\frac{1}{2}\right)$\hspace{1cm}&\hspace{1cm} Lorentz four-vector  \\
		$(1,0)\oplus(0,1)$\hspace{1cm}&\hspace{1cm} Maxwell electromagnetic fields \\
		$\left(\frac{3}{2},0\right)\oplus\left(0,\frac{3}{2}\right)$\hspace{1cm}&\hspace{1cm} Weinberg spinor (spin 3/2) \\
		$\left(1,1\right)$\hspace{1cm}&\hspace{1cm} Lorentz order-2 traceless symmetric tensor \\
		$\left(1,\frac{1}{2}\right)\oplus\left(\frac{1}{2},1\right)$\hspace{1cm}&\hspace{1cm} Rarita-Schwinger spinor (spin 3/2) \\
		$\left(2,0\right)\oplus\left(0,2\right)$\hspace{1cm}&\hspace{1cm} Einstein gravitational fields\\
		$\vdots$\hspace{1cm}&\hspace{1cm} \vdots \\
		\hline
	\end{tabular}
	\caption{The self-conjugate representations of $L_{p}$ with physical correspondence.}
	\label{tab:irreps_Lip}
\end{table}
Secondly, there are group elements of $L_p$ that make $k^\mu$ unchanged.
All of these elements construct a subgroup of $L_p$, which is known as the little group~\cite{Bargmann:1948ck}, denoted as $L_{p,k}$. 
The little groups of various standard momenta $k^\mu$ are shown in table~\ref{tab:little_group}. Except for the trivial case of vacuum momentum $k_{\mu}=\left(0,\mathbf{0}\right)$,
the other three cases correspond to three different little group chains of $L_p$ as shown in figure~\ref{fig:properlorentzgroupchain}. 
These different little group chains imply that one can choose different complete set of commuting operators (CSCO, see ref.~\cite{Chen:1985zzd}) to characterize Lorentz covariant wave functions. 
Here, we will discuss the case of little group chain containing SO(3). 
A Lorentz covariant wave function characterized by this little group chain is identified as spin wave function.
Due to the requirement of Lorentz covariance, under any rotation transformation $R\in\text{SO(3)}\subset L_{p}$, spin wave functions require~\cite{Weinberg:1964cn}
\begin{equation}
	D_{\alpha}^{~~\beta}(R)\, u^\sigma_\beta(\mathbf{k};s)=\, u^{\sigma'}_\alpha(\mathbf{k};s)\, D^{(s)\sigma}_{\sigma'}(R),\qquad D_{\alpha}^{~~\beta}(R)\, \bar{u}_\sigma^\alpha(\mathbf{k};s)=\, \bar{u}_{\sigma'}^\beta(\mathbf{k};s)~ D^{(s)\sigma'}_{\sigma}(R),
	\label{eq:spin_wave_function}
\end{equation} 
where $D^{(s)\sigma}_{\sigma'}(R)$ is Wigner-$D$ matrix in arbitrary IRREP of SO(3), labeled by spin $s$.
From eq.~\eqref{eq:spin_wave_function} and the definition of IRT in eq.~\eqref{eq:def_of_irten}, one can get $u^\sigma_\alpha(\mathbf{k};s)$ and $\bar{u}_\sigma^\alpha(\mathbf{k};s)$ which are IRTs of the little group SO(3).
The relevant two representations are $D_{\alpha}^{~~\beta}(L_{p,k})$ and $D^{(s)\sigma}_{\sigma'}(L_{p,k})$, respectively. 
According to this conclusion, one can obtain the specific form of the spin wave function in the standard momentum frame by calculating the IRT of the little group SO(3).
For example, we consider an IRREP of $L_p$, labeled by
\begin{equation}
	[\alpha]=(s_L,s_R)=(s_L,0)\otimes(0,s_R)\equiv[l]\otimes[r],\notag
\end{equation}
where $l=-s_L,-s_L+1,\cdots,s_L$ and $r=-s_R,-s_R+1,\cdots,s_R$ correspond to the IRREPs indices of SU$(2)_L$ and SU$(2)_R$, respectively. 
The corresponding covariant and contravariant spin wave functions with standard momentum are as follows,
\begin{align}
	u_{\alpha}^{\sigma}\left(\mathbf{k};s\right)\equiv u_{\alpha}^{\sigma}\left(\mathbf{k};(s_L,s_R),s\right)&=U_{\alpha}^{lr}~u_{lr}^{\sigma}\left(\mathbf{k};(s_L,s_R),s\right),\notag\\
	\bar{u}^{\alpha}_{\sigma}\left(\mathbf{k};s\right)\equiv\bar{u}^{\alpha}_{\sigma}\left(\mathbf{k};(s_L,s_R),s\right)&=U^{\alpha}_{lr}~\bar{u}^{lr}_{\sigma}\left(\mathbf{k};(s_L,s_R),s\right),
	\label{eq:spin_wave_func_irrep}
\end{align}
where $U_{\alpha}^{lr}$ and $U^{\alpha}_{lr}$ are two operations which flatten two indices $l~(-s_L\leq l\leq s_L)$ and $r~(-s_R\leq r\leq s_R)$ into one index $\alpha~\left(1\leq\alpha\leq(2s_L+1)(2s_R+1)\right)$ with the following explicit form, 
\begin{equation}
	U_{\alpha}^{lr}=\delta_{\,\alpha}^{~~\left[(l+s_L)(2s_R+1)+r+s_R+1\right]},\qquad
	U^{\alpha}_{lr}=\delta^{\,\alpha}_{~~\left[(l+s_L)(2s_R+1)+r+s_R+1\right]}.
	\label{eq:appd_index_map_alpha_to_lr}
\end{equation}
Then, eq.~\eqref{eq:spin_wave_function} can be written as
\begin{align}
	D_{l}^{(s_L)l'}(R)~D_{r}^{(s_R)r'}(R)\, u_{l'r'}^{\sigma}(\mathbf{k};(s_L,s_R),s)&=\, u_{lr}^{\sigma'}(\mathbf{k};(s_L,s_R),s)\, D^{(s)\sigma}_{\sigma'}(R),\notag\\
	D_{l'}^{(s_L)l}(R)~D_{r'}^{(s_R)r}(R)\, \bar{u}^{l'r'}_{\sigma}(\mathbf{k};(s_L,s_R),s)&=\, \bar{u}^{lr}_{\sigma'}(\mathbf{k};(s_L,s_R),s)\, D^{(s)\sigma'}_{\sigma}(R).
	\label{eq:spin_wave_function_chiral}
\end{align}
Comparing eq.~\eqref{eq:spin_wave_function_chiral} with eq.~\eqref{eq:trans_nature_CGCs}, one can immediately obtain $u_{lr}^{\sigma}(\mathbf{k};(s_L,s_R),s)$ and $\bar{u}^{lr}_{\sigma}(\mathbf{k};(s_L,s_R),s)$ which are just the Clebsch–Gordan coefficients (CGCs) of SU(2) (see also section XI of chapter V in ref.~\cite{weinberg_1995}),
\begin{equation}
	u_{lr}^{\sigma}(\mathbf{k};(s_L,s_R),s) = (C^s_{s_L s_R})^\sigma_{lr},\qquad
	\bar{u}^{lr}_{\sigma}(\mathbf{k};(s_L,s_R),s) = (C_s^{s_L s_R})_\sigma^{lr}.
	\label{eq:spin_wave_func_CGCs}
\end{equation}
Indeed, $u_{lr}^{\sigma}(\mathbf{k};(s_L,s_R),s)$ and $\bar{u}^{lr}_{\sigma}(\mathbf{k};(s_L,s_R),s)$ satisfy the orthonormal relation as shown in eq.~\eqref{eq:orthogonal_normalization}.

Futhermore, in any self-conjugate representation $[\alpha]=[s_L,s_R]~(s_L\neq s_R)$, a spin wave function can be straightforwardly written as,
\begin{align}
	u_{\alpha}^{\sigma}(\mathbf{k};s)&\equiv u_{\alpha}^{\sigma}(\mathbf{k};\chi,s)
	=\left\{\begin{array}{ll}
		\, \left(U_L\right)^{lr}_{\alpha} u^{\sigma}_{lr}(\mathbf{k};(s_L,s_R),s) ~~\text{with $\chi=(s_L,s_R)$} \\ \\ \, \left(U_R\right)^{lr}_{\alpha} u^{\sigma}_{lr}(\mathbf{k};(s_R,s_L),s) ~~\text{with $\chi=(s_R,s_L)$}
	\end{array}\right.,\notag\\
	\bar{u}^{\alpha}_{\sigma}(\mathbf{k};s)&\equiv\bar{u}^{\alpha}_{\sigma}(\mathbf{k};\chi,s)
	=\left\{\begin{array}{ll}
		\,\left(U_R\right)_{lr}^{\alpha} \bar{u}_{\sigma}^{lr}(\mathbf{k};(s_R,s_L),s) ~~\text{with $\chi=(s_L,s_R)$} \\
		\\
		\, \left(U_L\right)_{lr}^{\alpha} \bar{u}_{\sigma}^{lr}(\mathbf{k};(s_L,s_R),s) ~~\text{with $\chi=(s_R,s_L)$}
	\end{array}\right.,
	\label{eq:self_conj_rep_spin_wave_func}
\end{align} 
where $\left(U_{L/R}\right)^{lr}_{\alpha}$ and $\left(U_{L/R}\right)_{lr}^{\alpha}$ are four operations that flatten two indices $l~(-s_{L} \leq l \leq s_{L})$ and $r~(-s_{R} \leq r \leq s_{R})$ into one index $\alpha$ $(1 \leq \alpha \leq 2(2s_{L}+1) (2s_{R}+1))$ with the following explicit form,
\begin{align}
	\left(U_L\right)_{\alpha}^{lr}~&=~\delta_{\,\alpha}^{~~\left[(l+s_{L})(2s_{R}+1)+r+s_{R}+1\right]},\qquad \left(U_R\right)_{\alpha}^{lr}~=~\delta_{\,\alpha}^{~~\left[(3s_{L}+l+1)(2s_{R}+1)+r+s_{R}+1\right]},\notag\\
	\left(U_L\right)^{\alpha}_{lr}~&=~\delta^{\,\alpha}_{~~\left[(l+s_{L})(2s_{R}+1)+r+s_{R}+1\right]},\qquad
	\left(U_R\right)^{\alpha}_{lr}~=~\delta^{\,\alpha}_{~~\left[(3s_{L}+l+1)(2s_{R}+1)+r+s_{R}+1\right]}.
	\label{eq:appd_index_map_alpha_to_lr_rerep}
\end{align}
The corresponding orthonormal relation is as follows,
\begin{equation}
	\bar{u}^{\alpha}_{\sigma_1}(\mathbf{k};\chi_{_1}^*,s)\, u_{\alpha}^{\sigma_2}(\mathbf{k};\chi_{_2},s)\,=\,\delta_{\sigma_1}^{~~\sigma_2}\,\delta_{\chi_{_1}\chi_{_2}},
	\label{eq:self_conj_rep_oorthogonal_norm}
\end{equation}	
where $\chi_{_1}^*$ is the complex conjugate representation of $\chi_{_1}$.

Finally, by combining the Lorentz transformation matrix and the spin wave functions in standard momentum frame, one can obtain a spin wave function in any frame.
Furthermore, eq.~\eqref{eq:spin_wave_func_CGCs} implies that $u_{\alpha}^{\sigma}(\mathbf{p};\chi,s)$ can describe a particle with spin $s$ which satisfies the following triangular relation,
\begin{equation}
	|s_L-s_R|\leq s \leq s_L +s_R,
	\label{eq:select_rule_spin_wave_func}
\end{equation}
otherwise $u_{\alpha}^{\sigma}(\mathbf{p};\chi,s)$ must be zero. 
This means that there are many choices to express a spin wave function with spin $s$ in a Lorentz covariant way.
Actually, in the development of relativistic spin wave function, there are several formulations for particles with arbitrary spin $s$.
The most famous one is known as Rarita-Schwinger spin wave function~\cite{Rarita:1941mf}, which is based on the following self-conjugate representation,
\begin{align}
	\left[\alpha^{s}\right]&=\left\{
	\begin{array}{cl}
		\,\left(\frac{2 s+1}{4}, \frac{2 s-1}{4}\right)\oplus\left(\frac{2 s-1}{4}, \frac{2 s+1}{4}\right) & ~~\text{for half-integer }s\\
		\\
		\,\left(\frac{s}{2},\frac{s}{2}\right)& ~~\text{for integer }s
	\end{array}
	\right..
	\label{eq:massive_particle_rep}
\end{align}
These representations can be expressed by Lorentz four-vector indices for integer $s$ and an additional Dirac spinor index for half-integer $s$. 
Unfortunately, there are too many components of such spin wave function, including many irrelevant degrees of freedom. 
In 1948, Bargmann and Wigner proposed spin wave function which is based on the following self-conjugate reducible representation (REREP)~\cite{Bargmann:1948ck},
\begin{align}
	\left[\alpha^{s}\right]=\underbrace{\left[\frac12,0\right]\otimes\left[\frac12,0\right]\otimes\cdots\otimes\left[\frac12,0\right]}_{2s} ~~\text{for integer and half-integer }s.
	\label{eq:swf_BW_1948}
\end{align}
These representations can be expressed by Dirac spinor indices only, which make the expression more compact, but there are still many irrelevant degrees of freedom.  
In 1964, Weinberg proposed general covariant causal fields and suggested the following self-conjugate REREP~\cite{Weinberg:1964cn},
\begin{align}
	\left[\alpha^s\right]=\left(s,0\right)\oplus\left(0, s\right) ~~\text{for integer and half-intger }s.
\end{align}
These representations do not contain any redundant components, which make the expression elegant and concise. 
Subsequently, in order to further simplify the expression of scattering amplitude, only the left or right-handed spinor with spin-$\frac{1}{2}$ is retained as the basic quantity, which is abbreviated as 
$\lambda^I_\alpha/\tilde{\lambda}^{\dot{\alpha}}_{I}$ instead of $u_{l0}^{\sigma}\left(\mathbf{p};\left(\frac{1}{2},0\right),\frac12\right)/\bar{u}^{0r}_{\sigma}\left(\mathbf{p};\left(\frac{1}{2},0\right),\frac12\right)$ with $\alpha\mapsto l$, $\dot{\alpha}\mapsto r$ and $I\mapsto\sigma$ in our convention. 
Amplitudes including particles of higher spin can be expressed by tensor product and contraction of the basic quantity, which is called spinor-helicity formalism. %
For more details, one may refer to Nima's recent work on scattering amplitudes~\cite{Arkani-Hamed:2017jhn} and references therein.


\subsection{Irreducible tensors of $L_p$ and its little group SO(3)}
\label{sec:irten_Lp}

In this subsection, we will show how to derive the IRTs of $L_p$ and the IRTs of little group SO(3) based on spin wave functions which are introduced in subsection~\ref{sec:any_spin_wave_function}.

In the following discussion, we will use the Latin alphabet $a,~b,~c,\cdots$ to label Dirac spinor representation $\left(\frac{1}{2},0\right)\oplus\left(0,\frac{1}{2}\right)$; 
and use $a^{3},~b^{3},~c^{3},\cdots$ to label the Rarita-Schwinger spinor representation $\left(1, \frac{1}{2}\right)\oplus\left(\frac{1}{2}, 1\right)$; 
and use the Greek alphabet $\mu,~\nu,~\rho~,\cdots$ to label the Lorentz four-vector representation $\left(\frac{1}{2},\frac{1}{2}\right)$; 
and use $\mu^{2},~\nu^{2},~\rho^{2},\cdots$ to label the representation $\left(1,1\right)$;
and see table~\ref{tab:symbol_conventions} for other cases.
In general, we will use $\alpha,~\beta$ to label any representation of $L_p$. 
We will focus on the self-conjugate representations.
\begin{table}
	\centering
	\begin{tabular}{c|c|c}
		\hline
		Rep. of $L_p$ & Notations of Rep. & Indices of Rep. \\
		\hline
		~~$(s,0)~(2s\in\mathbbm{N}^*)$~~&~~$[l]$~~&~~$l$~~\\
		~~$(0,s)~(2s\in\mathbbm{N}^*)$~~&~~$[r]$~~&~~$r$~~\\
		~~$\left(\frac{2 s+1}{4}, \frac{2 s-1}{4}\right)\oplus\left(\frac{2 s-1}{4}, \frac{2 s+1}{4}\right)~\left(s\in\mathbbm{N}+\frac{1}{2}\right)$~~ & ~~$[a^{2s}],~[b^{2s}],\cdots$~~ & ~~$a^{2s},~b^{2s},\cdots$~~ \\
		~~$\left(\frac{2 s+1}{4}, \frac{2 s-1}{4}\right)~\left(s\in\mathbbm{N}+\frac{1}{2}\right)$~~ & ~~$[a_L^{2s}],~[b_L^{2s}],\cdots$~~ & ~~$a_L^{2s},~b_L^{2s},\cdots$~~ \\
		~~$\left(\frac{2 s-1}{4}, \frac{2 s+1}{4}\right)~\left(s\in\mathbbm{N}+\frac{1}{2}\right)$~~ & ~~$[a_R^{2s}],~[b_R^{2s}],\cdots$~~ & ~~$a_R^{2s},~b_R^{2s},\cdots$~~ \\
		$\left(\frac{s}{2},\frac{s}{2}\right)~(s\in\mathbbm{N})$ ~~ & ~~$[\mu^{s}],~[\nu^{s}],\cdots$~~ & $\mu^{s},~\nu^{s},\cdots$~~ \\
		~~any other representation~~ & ~~$[\zeta],~[\xi],\cdots$~~ & ~~$\zeta,~\xi,\cdots$~~ \\
		\hline
	\end{tabular}
\caption{The first column is the representations of the $L_p$. The second column is the simplified notations for these representations. The third column is the index for each representation space.}
	\label{tab:symbol_conventions}
\end{table}
An order-3 IRT of $L_p$, denoted as $T^{\alpha_1\alpha_2}_\beta$, is a projection operator from the representation space $[\alpha_1]\otimes[\alpha_2]$ to its representation subspace $[\beta]$. 
Thus, one can obtain an order-3 IRT $T^{\alpha_1\alpha_2}_\beta$ by using spin wave functions as follows,
\begin{equation}
	T^{\alpha_1\alpha_2}_\beta =\,\sum_{\chi,s}\, u^{\sigma}_\beta(\mathbf{k};\chi,s)\,\bar{u}_{\sigma}^{\alpha_1\alpha_2}(\mathbf{k};\chi^*,s),
	\label{eq:irten_order_3_CGCs}
\end{equation}
where the summation of $\chi$ spans all possible IRREPs belonging to both the representation $[\beta]$ and the direct product representation $[\alpha_1]\otimes[\alpha_2]$;
the summation of $s$ spans all possible spins in $\chi$ from the selection rule as shown in eq.~\eqref{eq:select_rule_spin_wave_func}.
The specific form of spin wave functions with two Lorentz covariant / contravariant indices are as follows,
\begin{align}
	u^{\sigma}_{\alpha_1\alpha_2}\left(\mathbf{k};\chi,s\right) \,=\, \sum_{s_1,s_2} \,&\sqrt{(2s_1+1)(2s_2+1)(2s_L+1)(s_R+1)}\, \left\{\begin{array}{ccc}
		s_{1L} & s_{1R} & s_1 \\
		s_{2L} & s_{2R} & s_2 \\
		s_{L} &  s_{R} &   s \\
	\end{array}\right\}\notag\\
	&\times\, \left(C^s_{s_1s_2}\right)^\sigma_{\sigma_1\sigma_2}\, u_{\alpha_1}^{\sigma_1}\left(\mathbf{k};\chi_{_1},s_1\right)\,u_{\alpha_2}^{\sigma_2}\left(\mathbf{k};\chi_{_2},s_2\right),\notag\\
	\bar{u}_{\sigma}^{\alpha_1\alpha_2}\left(\mathbf{k};\chi^*,s\right) \,=\, \sum_{s_1,s_2} \,&\sqrt{(2s_1+1)(2s_2+1)(2s_L+1)(s_R+1)}\, \left\{\begin{array}{ccc}
		s_{1L} & s_{1R} & s_1 \\
		s_{2L} & s_{2R} & s_2 \\
		s_{L} &  s_{R} &   s \\
	\end{array}\right\}\notag\\
	&\times\, \left(C_s^{s_1s_2}\right)_\sigma^{\sigma_1\sigma_2}\, \bar{u}^{\alpha_1}_{\sigma_1}\left(\mathbf{k};\chi_{_1}^*,s_1\right)\,\bar{u}^{\alpha_2}_{\sigma_2}\left(\mathbf{k};\chi_{_2}^*,s_2\right),
	\label{eq:spin_wave_func_direct_product}
\end{align}
with $\chi=(s_L,s_R)$, $\chi_{_1}=(s_{1L},s_{1R})$ and $\chi_{_2}=(s_{2L},s_{2R})$,
where the summation of $s_{1/2}$ spans all possible spins in $\chi_{_{1/2}}^*$ from the selection rule as shown in eq.~\eqref{eq:select_rule_spin_wave_func}; $\chi$ is arbitrary IRREP belonging to $[\alpha_1]\otimes[\alpha_2]$; $\chi_{_{1/2}}$ is arbitrary IRREP belonging to $[\alpha_{1/2}]$;
$\{\cdots\}$ denotes Wigner-9$j$ symbol.
By using order-3 IRTs, any order-$n$ IRTs can be obtained recursively. 
For convenience, we give some examples in appendix~\ref{appd:ex_irten_Lp}.

For IRTs of little group SO(3), according to the definition of generators $A_i$ and $B_i~(i=1,2,3)$ as given by eq.~\eqref{eq:def_A_B}, there is no difference between the left-handed representation $(s_L,0)$ and the right-handed representation $(0,s_R)$ for rotation transformation.
In addition, the spin quantum number $s$ is an invariant quantity. Thus, one can get arbitrary order-3 IRT of little group SO(3) by removing the restriction on type of IRREPs ($\chi\mapsto\chi^*$) and the summation of spins $s$ in eq.~\eqref{eq:irten_order_3_CGCs},
\begin{equation}
	P^{\alpha_1\alpha_2}_\beta(\mathbf{p};\chi_{_1},\chi_{_2},s) = u^{\sigma}_\beta(\mathbf{p};\chi_{_1},s)\,\bar{u}_{\sigma}^{\alpha_1\alpha_2}(\mathbf{p};\chi_{_2}^*,s),
	\label{eq:irten_little_group_so3}
\end{equation}
where $\chi_{_1}$ and $\chi_{_2}$ are arbitrary IRREPs belonging to the representation $[\beta]$ and $[\alpha_1]\otimes[\alpha_2]$, respectively.
By employing eq.~\eqref{eq:spin_wave_func_direct_product}, it is more convenient to use the following definition,
\begin{equation}
	P^{\alpha_1\alpha_2}_\beta(\mathbf{p};\chi,s;\chi_{_1},s_1;\chi_{_2},s_2) = \left(C_s^{s_1s_2}\right)_\sigma^{\sigma_1\sigma_2}\,u^{\sigma}_\beta(\mathbf{p};\chi,s)\, \bar{u}^{\alpha_1}_{\sigma_1}\left(\mathbf{p};\chi_{_1}^*,s_1\right)\,\bar{u}^{\alpha_2}_{\sigma_2}\left(\mathbf{p};\chi_{_2}^*,s_2\right).
	\label{eq:spin_proj_ten_three_chi}
\end{equation}
IRTs of little group SO(3) are always called Lorentz covariant spin projection tensors.
It is worth noting that the components of different spin states of arbitrary IRREP $(s_L,s_R)$ will mix with each other after a Lorentz boost transformation, unless there is only one possible spin value in the IRREP.\footnote{This is only true for IRREP $(s_L,s_R)$ with $s_L=0$ or $s_R=0$, i.e., the Weinberg spin wave function as discussed in section~\ref{sec:any_spin_wave_function}.}
Therefore, spin projection tensors are depend on the momentum $\mathbf{p}$, and they are invariant under rotation.
To gain a more intuitive understanding of IRTs of the little group SO(3), we provide examples in appendix~\ref{appd:ex_irten_so3}.


\subsection{Spin wave function of massless particle}
\label{sec:massless_spin_wave_func}

In subsection~\ref{sec:any_spin_wave_function}, we did not mention the difference of the spin wave functions of particles with and without mass. 
In this subsection, we will focus on this point.

As given in table~\ref{tab:little_group}, the little group of a massive particle is SO(3).
The corresponding group element $h_\mathbf{p}$ (eq.~\eqref{eq:move_wave_function}) must belong to $L_p/\text{SO(3)}$ which corresponds to a pure-boost Lorentz transformation, and can be written as follows,
\begin{equation}
	L_p/\text{SO(3)}\,\ni\,h_\mathbf{p}\,=\,R_{\hat{\mathbf{p}}}\,\cdot\,B_{|\mathbf{p}|}\,\cdot\,R^{-1}_{\hat{\mathbf{p}}},
	\label{eq:convention_massive_h_p}
\end{equation}
where $R_{\hat{\mathbf{p}}}$ is a space-rotation transformation from $z$-axis to the direction of $\mathbf{p}$, denoted as $\hat{\mathbf{p}}$; 
and $B_{|\mathbf{p}|}$ is a pure-boost transformation along $z$-axis which makes the energy of a massive particle at rest from $m$ to $\omega_\mathbf{p}$. 
Thus, for a massive particle, the general form of spin wave function has been discussed in subsection~\ref{sec:any_spin_wave_function}. 
We will adopt Rarita-Schwinger spin wave function (eq.~\eqref{eq:massive_particle_rep}) for a massive particle by default in the remaining of this paper.

For massless particles, since the little group is ISO(2), the corresponding group element $\tilde{h}_\mathbf{p}$ must belong to $L_p/\text{ISO(2)}$, which will not correspond to pure-boost Lorentz transformation,
and can be written as follows (see section~V of chapter~II in ref.~\cite{weinberg_1995} for details),
\begin{equation}
	L_p/\text{ISO(2)}\,\ni\,\tilde{h}_\mathbf{p}\,=\,R_{\hat{\mathbf{p}}}\,\cdot\,B_{|\mathbf{p}|}.
	\label{eq:convention_massless_h_p}
\end{equation}
Then, similar with eq.~\eqref{eq:move_wave_function}, one has
\begin{equation}
	u^{\sigma}_{\alpha}\left(\mathbf{p};\chi,\tilde{s}\right) =\, D_{\alpha}^{~~\beta}\left(\tilde{h}_{\mathbf{p}}\right)\,u^{\sigma}_{\beta}\left(\mathbf{k};\chi,\tilde{s}\right),\qquad \bar{u}_{\sigma}^{\alpha}\left(\mathbf{p};\chi,\tilde{s}\right) =\, D_{\beta}^{~~\alpha}\left(\tilde{h}_{\mathbf{p}}^{-1}\right)\,\bar{u}_{\sigma}^{\beta}\left(\mathbf{k};\chi,\tilde{s}\right),
	\label{eq:def_helicity_wave_func}
\end{equation}
where $\tilde{s}$ is eigenvalue of Casimir operator $C_{\text{ISO(2)}}$ of ISO(2), which labels IRREPs of ISO(2).
Since the goal of this paper is to construct Lorentz covariant partial wave amplitude, according to eq.~\eqref{eq:any_spin_general_LS_1_rest}, the pure-spin part is indispensable.
We require a definite spin $s$ for $u^{\sigma}_{\alpha}\left(\mathbf{k};\chi,\tilde{s}\right)$ / $\bar{u}_{\sigma}^{\alpha}\left(\mathbf{k};\chi,\tilde{s}\right)$, then it is rewritten as $u^{\sigma}_{\alpha}\left(\mathbf{k};\chi,s,\tilde{s}\right)$ / $\bar{u}_{\sigma}^{\alpha}\left(\mathbf{k};\chi,s,\tilde{s}\right)$. 
In other words, it is an eigenstate of the Casimir operator $C_{\text{SO(3)}}$ of SO(3).
Thus, $u^{\sigma}_{\alpha}\left(\mathbf{k};\chi,s,\tilde{s}\right)$ / $\bar{u}_{\sigma}^{\alpha}\left(\mathbf{k};\chi,s,\tilde{s}\right)$ should be an eigenstate of both $C_{\text{ISO(2)}}$ and $C_{\text{SO(3)}}$, as well as the $J_3$ which is the Casimir operator of SO(2).
Correspondingly, from the commutation relations as shown in table~\ref{tab:little_group}, one can obtain
\begin{equation}
	C_{\text{ISO(2)}} =\, \left(K_1-J_2\right)^2+\left(K_2+J_1\right)^2,\qquad
	C_{\text{SO(3)}} =\, J_1^2+J_2^2+J_3^2.
	\label{eq:casimir_so3_iso2}
\end{equation}
The eigen-equations of $u^{\sigma}_{\alpha}\left(\mathbf{k};\chi,s,\tilde{s}\right)$ can be expressed as follows,
\begin{align}
	\left[\,J_3\,\right]_{\alpha}^{~~\beta}\,u^{\sigma}_{\beta}\left(\mathbf{k};\chi,s,\tilde{s}\right) &=\, \sigma\,u^{\sigma}_{\alpha}\left(\mathbf{k};\chi,s,\tilde{s}\right),\notag\\
	\left[C_{\text{ISO(2)}}\right]_{\alpha}^{~~\beta}\,u^{\sigma}_{\beta}\left(\mathbf{k};\chi,s,\tilde{s}\right) &=\, \tilde{s}\,u^{\sigma}_{\alpha}\left(\mathbf{k};\chi,s,\tilde{s}\right),\notag\\
	\left[C_{\text{SO(3)}}\right]_{\alpha}^{~~\beta}\,u^{\sigma}_{\beta}\left(\mathbf{k};\chi,s,\tilde{s}\right) &=\, s(s+1)\,u^{\sigma}_{\alpha}\left(\mathbf{k};\chi,s,\tilde{s}\right).
	\label{eq:spin_helicity_wave_function}
\end{align}
The eigen-equations of $\bar{u}_{\sigma}^{\alpha}\left(\mathbf{k};\chi,s,\tilde{s}\right)$ are similar.

Then, by combining eqs.~\eqref{eq:casimir_so3_iso2} and \eqref{eq:spin_helicity_wave_function}, the spin wave function for a massless particle with spin $s$ and helicity $\sigma=\pm s$ must be described in the following self-conjugate representation,
\begin{equation}
	\left[\zeta^s\right]=\left(s,0\right)\oplus\left(0,s\right),
	\label{eq:self_conj_rep_spin_helicity}
\end{equation}
and one also has $\tilde{s}=0$ in these representations. 
For simplicity, we can rewrite $u^{\sigma}_{\zeta^s}\left(\mathbf{k};\chi,s,\tilde{s}\right)$ / $\bar{u}_{\sigma}^{\zeta^s}\left(\mathbf{k};\chi,s,\tilde{s}\right)$ as $u^{\sigma}_{\zeta^s}\left(\mathbf{k};\chi,s\right)$ / $\bar{u}_{\sigma}^{\zeta^s}\left(\mathbf{k};\chi,s\right)$, which has the same arguments as the spin wave function of massive particles.
One will find the spin wave function of massless particles can be projected from the corresponding spin wave function of massive particles.
Such projection operator is named as helicity projection tensor.
Now, we will explain how to construct these helicity projection tensors.
%


Above all, we note that the REREP $[s,0]$ is included in the following  decompositions (with the convention as given in table~\ref{tab:symbol_conventions}),
\begin{align}
	\left[\alpha^{s}\right]\otimes\left[\,\nu^{\bar{s}}\,\right]=\left[\zeta^s\right]\oplus\cdots,
\end{align}
where $\bar{s}=\lfloor s\rfloor$ (rounding down). 
This decomposition indicates the existence of IRTs as follows,
\begin{equation}
	\left[\alpha^{s}\right]\otimes\left[\,\nu^{\bar{s}}\,\right]~\mapsto~\left(s,0\right) ~:~ T^{\alpha^{s}\nu^{\bar{s}}}_{l},\qquad
	\left[\alpha^{s}\right]\otimes\left[\,\nu^{\bar{s}}\,\right]~\mapsto~\left(0,s\right) ~:~ T^{\alpha^{s}\nu^{\bar{s}}}_{r},
\end{equation}
where the index $\nu^{\bar{s}}$ can be replaced by Lorentz four-vector indices by using IRT $T_{\nu^{\bar{s}}}^{\nu_1\nu_2\cdots\nu_{\bar{s}}}$ as follows,
\begin{equation}
	T^{\alpha^s\nu_1\nu_2\cdots\nu_{\bar{s}}}_{l} = T^{\alpha^s\nu^{\bar{s}}}_{l} ~ T_{\nu^{\bar{s}}}^{\nu_1\nu_2\cdots\nu_{\bar{s}}},\qquad
	T^{\alpha^s\nu_1\nu_2\cdots\nu_{\bar{s}}}_{r} = T^{\alpha^s\nu^{\bar{s}}}_{r} ~ T_{\nu^{\bar{s}}}^{\nu_1\nu_2\cdots\nu_{\bar{s}}}.
	\label{eq:irten_Lp_any_spin_massless}
\end{equation}
One can get these IRTs from eq.~\eqref{eq:irten_order_3_CGCs}. 

Then, since the standard momentum of a massless particle, $k_\mu=(|\mathbf{k}|,0,0,|\mathbf{k}|)$, is covariant under Lorentz transformation and invariant under transformation of little group ISO(2), the contractions between $k_\mu$ and IRTs in eq.~\eqref{eq:irten_Lp_any_spin_massless} have the same properties.
In other words, one will have the following helicity projection tensors,
\begin{equation}
	P^{\alpha^s}_{l}\left(\mathbf{k};s\right) = T^{\alpha^s\nu_1\nu_2\cdots\nu_{\bar{s}}}_{l}\,k_{\nu_1}\,k_{\nu_2}\cdots k_{\nu_{\bar{s}}},\qquad
	P^{\alpha^s}_{r}\left(\mathbf{k};s\right) = T^{\alpha^s\nu_1\nu_2\cdots\nu_{\bar{s}}}_{r}\,k_{\nu_1}\,k_{\nu_2}\cdots k_{\nu_{\bar{s}}}.
\end{equation}
With the above helicity projection tensors, one can convert the massive spin wave function $u^{\sigma}_{\alpha^s}\left(\mathbf{k};\chi,s\right)$ / $\bar{u}_{\sigma}^{\alpha^s}\left(\mathbf{k};\chi,s\right)$ into the massless spin wave function $u^{\sigma}_{\zeta^s}\left(\mathbf{k};\chi,s\right)$ / $\bar{u}_{\sigma}^{\zeta^s}\left(\mathbf{k};\chi,s\right)$, the relations are as follows,
\begin{align}
	u_{\zeta^s}^{\sigma}\left(\mathbf{k};(s,0),s\right) &=\,\lim_{|\mathbf{p}|\to\infty}\,N\left(|\mathbf{p}|\right)\,\left(U_L\right)_{\zeta^s}^{l0}\, P^{\alpha^s}_{l}\left(\mathbf{k};s\right) \, D_{\alpha^s}^{~~\beta^s}\left(B_{|\mathbf{p}|}\right)\,u_{\beta^s}^{\sigma}\left(\mathbf{0};\chi_{_L},s\right),\notag\\
	u_{\zeta^s}^{\sigma}\left(\mathbf{k};(0,s),s\right) &=\,\lim_{|\mathbf{p}|\to\infty}\,N\left(|\mathbf{p}|\right)\,\left(U_R\right)_{\zeta^s}^{0r}\, P^{\alpha^s}_{r}\left(\mathbf{k};s\right) \, D_{\alpha^s}^{~~\beta^s}\left(B_{|\mathbf{p}|}\right)\,u_{\beta^s}^{\sigma}\left(\mathbf{0};\chi_{_R},s\right),\notag\\
	\bar{u}^{\zeta^s}_{\sigma}\left(\mathbf{k};(s,0)^*,s\right) &=\,\lim_{|\mathbf{p}|\to\infty}\,N\left(|\mathbf{p}|\right)\,\left(U_L\right)^{\zeta^s}_{l0}~ P_{\alpha^s}^{l}\left(\mathbf{k};s\right)\, D^{~~\alpha^s}_{\beta^s}\left(B_{|\mathbf{p}|}^{-1}\right)\,\bar{u}^{\beta^s}_{\sigma}\left(\mathbf{0};\chi_{_L}^*,s\right),\notag\\
	\bar{u}^{\zeta^s}_{\sigma}\left(\mathbf{k};(0,s)^*,s\right) &=\,\lim_{|\mathbf{p}|\to\infty}\,N\left(|\mathbf{p}|\right)\,\left(U_R\right)^{\zeta^s}_{0r}\, P_{\alpha^s}^{r}\left(\mathbf{k};s\right)\, D^{~~\alpha^s}_{\beta^s}\left(B_{|\mathbf{p}|}^{-1}\right)\,\bar{u}^{\beta^s}_{\sigma}\left(\mathbf{0};\chi_{_R}^*,s\right),
	\label{eq:spin_wave_func_massless_half/integer}
\end{align}
where $N\left(|\mathbf{p}|\right)=1~\left(|\mathbf{p}|^{-1/2}\right)$ is a normalization factor for bosons (fermions);
$U_{L/R}$ is defined in eq.~\eqref{eq:appd_index_map_alpha_to_lr_rerep};
$B_{|\mathbf{p}|}$ is defined in eq.~\eqref{eq:convention_massive_h_p};
$\mathbf{0}$ denotes the spatial part of standard momentum of little group SO(3);
$\chi_{_{L/R}}=\left[\alpha^s\right]$ for arbitray self-conjugate IRREP, otherwise corresponds to the left/right-handed IRREP belonging to $\left[\alpha^s\right]$. 
It is worth pointing out that the limit in eqs.~\eqref{eq:spin_wave_func_massless_half/integer} is to remove the non-transverse polarized components in the spin wave functions, and has no effect on the remaining two transverse polarized components. 
Thus, after some derivation, the massless spin wave functions can be reduced as follows,
\begin{equation}
	\begin{array}{ll}
		u_{\zeta^s}^{\sigma}\left(\mathbf{k};(s,0),s\right)=\delta_{\zeta^s}^{1}\,\delta_{-s}^\sigma,\qquad&
		u_{\zeta^s}^{\sigma}\left(\mathbf{k};(0,s),s\right)=\delta_{\zeta^s}^{2(2s+1)}\,\delta_{s}^\sigma,\\
		\vspace{-3mm}&\\
		\bar{u}^{\zeta^s}_{\sigma}\left(\mathbf{k};(s,0)^*,s\right)=\delta^{\zeta^s}_{1}\,\delta^{-s}_\sigma,\qquad&
		\bar{u}^{\zeta^s}_{\sigma}\left(\mathbf{k};(0,s)^*,s\right)=\delta^{\zeta^s}_{2(2s+1)}\,\delta^{s}_\sigma.
	\end{array}
	\label{eq:spin_wave_func_massless_explicit}
\end{equation}

Finally, we need to separate $u^{\sigma}_{\zeta^s}\left(\mathbf{p};\chi,s\right)$ / $\bar{u}_{\sigma}^{\zeta^s}\left(\mathbf{p};\chi,s\right)$ into the pure orbital part and the pure spin part.
Thus, according to eqs.~\eqref{eq:any_spin_general_LS_1_rest} and \eqref{eq:convention_massless_h_p}, for the covariant/contravariant spin wave function, we will adopt the following separation,
\begin{align}
	u^{\sigma}_{\zeta^s}(\mathbf{p};\chi,s) ~&=~ D_{\zeta^s}^{~~\zeta^s_1}\left(\tilde{h}_{\mathbf{p}}\right)~u^{\sigma}_{\zeta^s_1}(\mathbf{k};\chi,s)
	~=~\underbrace{D_{\zeta^s}^{~~\zeta^s_1}\left(R_{\hat{\mathbf{p}}}\right)\,D_{\zeta^s_1}^{~~\zeta^s_2}\left(B_{|\mathbf{p}|}\right)}_{\text{pure orbital part}}~\underbrace{u^{\sigma}_{\zeta^s_2}(\mathbf{k};\chi,s)}_{\text{pure spin part}},\notag\\
	\bar{u}_{\sigma}^{\zeta^s}(\mathbf{p};\chi,s) ~&=~ D_{\zeta^s_1}^{~~\zeta^s}\left(\tilde{h}^{-1}_{\mathbf{p}}\right)~\bar{u}_{\sigma}^{\zeta^s_1}(\mathbf{k};\chi,s)
	~=~\underbrace{D_{\zeta^s_1}^{~~\zeta^s}\left(R^{-1}_{\hat{\mathbf{p}}}\right)\,D_{\zeta^s_2}^{~~\zeta^s_1}\left(B_{|\mathbf{p}|}^{-1}\right)}_{\text{pure orbital part}}~\underbrace{\bar{u}_{\sigma}^{\zeta^s_2}(\mathbf{k};\chi,s)}_{\text{pure spin part}}.
	\label{eq:spin_wave_func_standard_to_another}
\end{align}

In addition, from table~\ref{tab:irreps_Lip}, the self-conjugate representation $\left[\zeta^1\right]=(1,0)\oplus(0,1)$ carries Maxwell fields, which is the adjoint representation of $L_p$.
The dimension of this representation is 6, with three vectors and three axial vectors that respectively correspond to the electric and magnetic field components.
The self-conjugate representation $\left[\zeta^2\right]=(2,0)\oplus(0,2)$ carries Einstein gravitational field.
The dimension of this representation is 10, corresponding to the ten components of Riemann curvature tensor.
Therefore, describing the spin wave function of a massless particle through the self-conjugate representation $\left[\zeta^s\right]$ and field strength tensor is equivalent.
It automatically satisfies gauge invariance when the scattering process involves gauge bosons.


\subsection{Explicit form of covariant $L$-$S$ scheme and its relation with helicity scheme}
\label{sec:irten_exact}

In this subsection, we will construct Lorentz covariant partial wave amplitude by using IRTs of little group SO(3).
In subsection~\ref{sec:covariant_General}, we have shown
\begin{align} 
	\Gamma_{\alpha_1}^{\alpha_2\alpha_3}(\mathbf{k}_1,\mathbf{p}^*_2,\mathbf{p}^*_3;L,S)=\,P^{\alpha^L\alpha^S}_{\alpha_1}(\mathbf{k}_1;s_1,L,S)\,P^{\alpha_2\alpha_3}_{\alpha^S}(\mathbf{k}_1;S,s_2,s_3)\,\tilde{t}^{(L)}_{\alpha^L}(\mathbf{k}_1,\mathbf{p}^*_2-\mathbf{p}^*_3),
	\label{eq:general_LS_expand_1_rest_Gamma}
\end{align} 
with
\begin{align}
	\tilde{t}^{(L)}_{\alpha^L}(\mathbf{k}_1,\mathbf{p}^*_2-\mathbf{p}^*_3)&\equiv P^{\beta_1\cdots\beta_L}_{\alpha^L}(\mathbf{k}_1;L)~(p^*_2-p^*_3)_{\beta_1}\cdots (p^*_2-p^*_3)_{\beta_L},\notag\\
	P^{\beta_1\cdots\beta_L}_{\alpha^L}(\mathbf{k}_1;L)&= P^{\beta_1\alpha^{L-1}}_{\alpha^L}(\mathbf{k}_1;L,1,L-1)~ P^{\beta_2\cdots\beta_L}_{\alpha^{L-1}}(\mathbf{k}_1;L-1),
	\label{eq:t_tilde}
\end{align}
where $(p^*_2-p^*_3)_{\beta}=\left(U^{-1}\right)_\beta^{~~\mu}(p^*_2-p^*_3)_{\mu}$ with $\left(U^{-1}\right)_\beta^{~~\mu}$ is defined in eq.~\eqref{eq:appd_spacetime_chiral};
and the general form of order-3 spin projection tensor is as follows,
\begin{equation}
	P_{\alpha_1}^{\alpha_2\,\alpha_3}\left(\mathbf{k}_1;j_1,j_2,j_3\right) ~=~ \sum_{\chi_{_1},\chi_{_2},\chi_{_3}}~C_{\chi_{_1}\chi_{_2}\chi_{_3}}~
	P_{\alpha_1}^{\alpha_2\,\alpha_3}(\mathbf{k}_1;\chi_{_1},j_1;\chi_{_2},j_2;\chi_{_3},j_3),
	\label{eq:spin_proj_ten_JLS}
\end{equation}
where $P_{\alpha_1}^{\alpha_2\,\alpha_3}(\mathbf{k}_1;\chi_{_1},j_1;\chi_{_2},j_2;\chi_{_3},j_3)$ is defined in eq.~\eqref{eq:spin_proj_ten_three_chi}.
The $C_{\chi_{_1}\chi_{_2}\chi_{_3}}$s are indeterminate coefficients, which represent the degrees of freedom of the Lorentz covariant coupling structure.\footnote{ It is worth pointing out that according to Wigner-Eckart theorem, the form of the partial wave amplitude is only related to the values of $L$ and $S$ and is independent of $C_{\chi_{_1}\chi_{_2}\chi_{_3}}$s.
	In fact, any change in $C_{\chi_{_1}\chi_{_2}\chi_{_3}}$s can be absorbed into the definition of coupling constants. If one further requires coupling structure with conserved quantities such as $\mathcal{P}$-parity and $\mathcal{C}$-parity, $C_{\chi_{_1}\chi_{_2}\chi_{_3}}$s will be limited by additional conditions. An example of this can be seen in subsection~\ref{sec:ex_12_121}.}

\begin{figure}[h]
	\centering
	\includegraphics[width=\linewidth]{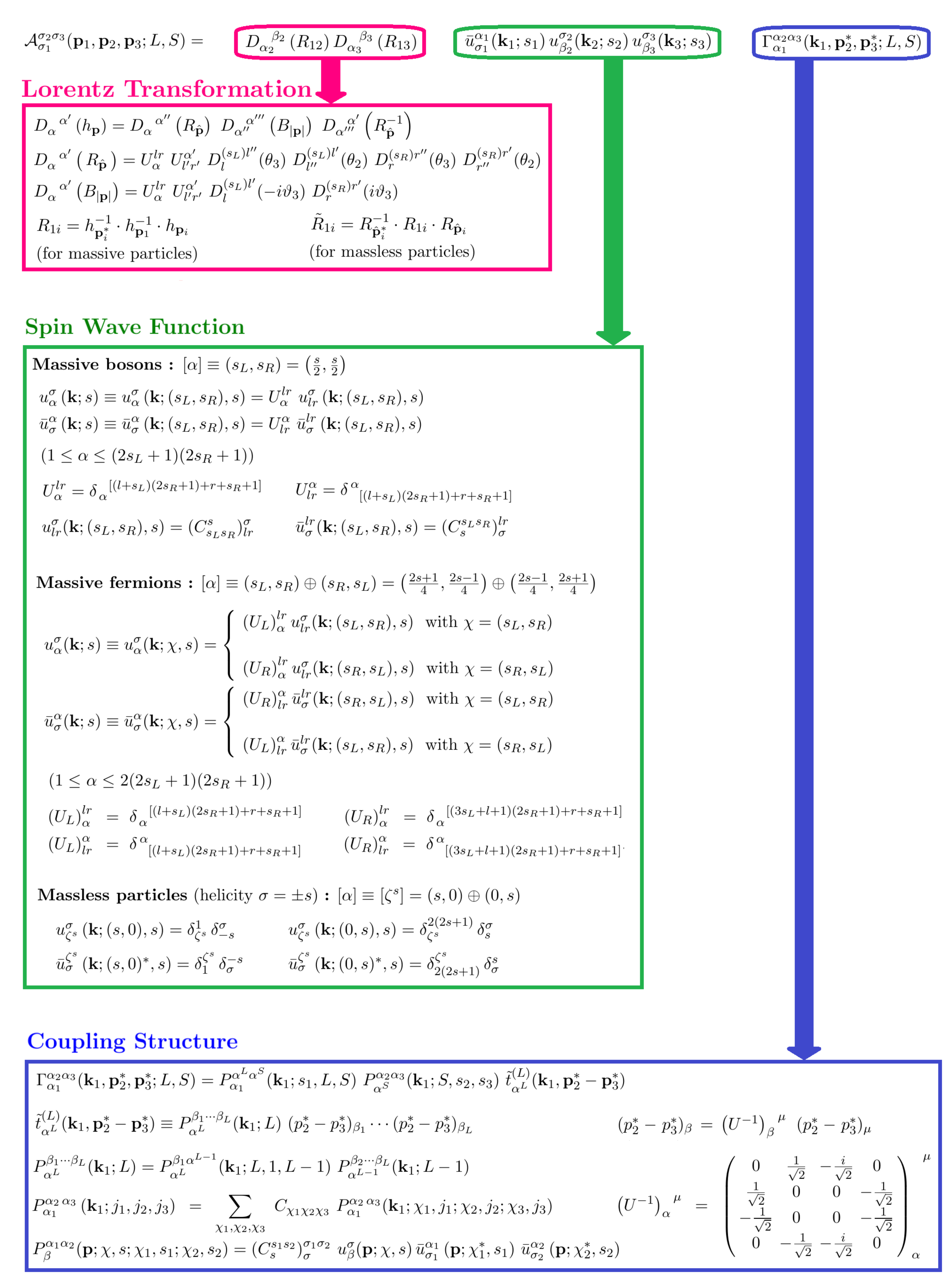}
	\caption{A panoramic view of deriving partial wave amplitude by using covariant $L$-$S$ scheme.}
	\label{fig:flowchart}
\end{figure}

By adopting $[\alpha_i]=\left[\alpha^{s_i}\right]$ as shown in eq.~\eqref{eq:massive_particle_rep}, one can easily obtain the exact expression of ${\cal A}_{\sigma_1}^{\sigma_2\sigma_3}(\mathbf{p}_1,\mathbf{p}_2,\mathbf{p}_3;L,S)$ for a process that only involves massive particles based on eq.~\eqref{eq:general_LS_expand_1_motion}.
For a process that involves massless particles, the explicit form is basically the same as that in eq.~\eqref{eq:general_LS_expand_1_motion}. 
The only difference is that, for massless particle, one needs to replace the spin wave function $u^{\sigma}_{\alpha_i}(\mathbf{k}_i;s)$ / $\bar{u}_{\sigma}^{\alpha_i}(\mathbf{k}_i;s)$ and the Lorentz transformation $R_{1i}$ in eq.~\eqref{eq:general_LS_expand_1_motion} with  $u^\sigma_{\zeta^{s_i}}\left(\mathbf{k}_i;\chi,s\right)$ / $\bar{u}_\sigma^{\zeta^{s_i}}\left(\mathbf{k}_i;\chi,s\right)$ and $\tilde{R}_{1i}=R_{\hat{\mathbf{p}}_i^*}^{-1} \cdot R_{1i} \cdot R_{\hat{\mathbf{p}}_i}$, respectively.
As a summary, the covariant $L$-$S$ scheme provides the general steps for calculating partial wave amplitude 
${\cal A}_{\sigma_1}^{\sigma_2\,\sigma_3}(\mathbf{p}_1,\mathbf{p}_2,\mathbf{p}_3;L,S)$, which are summarized in figure~\ref{fig:flowchart}.

In addition, it is useful to show the relation between the partial wave amplitude obtained by covariant $L$-$S$ scheme and the helicity formalism~\cite{Jacob:1959at}.
By combining eqs.~\eqref{eq:general_LS_expand_1_rest} and \eqref{eq:coupling_structure_Gamma_1_rest}, the partial wave amplitude at the rest frame of the initial particle can be written as follows,
\begin{align}
	{\cal A}_{\sigma_1}^{\sigma_2\sigma_3}(\mathbf{k}_1,\mathbf{p}^*_2,\mathbf{p}^*_3;L,S)=& \,P^{\alpha^L\alpha^S}_{\alpha_1}(\mathbf{k}_1;s_1,L,S)\,P^{\alpha_2\alpha_3}_{\alpha^S}(\mathbf{k}_1;S,s_2,s_3)\,\tilde{t}^{(L)}_{\alpha^L}(\mathbf{k}_1,\mathbf{p}^*_2-\mathbf{p}^*_3)\notag\\
	&\,\times
	\bar{u}_{\sigma_1}^{\alpha_1}(\mathbf{k}_1;s_1)\,u^{\sigma_2}_{\alpha_2}(\mathbf{k}_2;s_2)\,u^{\sigma_3}_{\alpha_3}(\mathbf{k}_3;s_3),
	\label{eq:general_LS_explicity_1_rest}
\end{align}
where three polarization indices $\sigma_i~(i=1,2,3)$ refer to the polarization component along the $z$-axis.\footnote{As shown in eq.~\eqref{eq:spin_wave_func_standard_to_another}, we also use helicity wave function for massless particle in covariant $L$-$S$ scheme. Thus, the relation between these two schemes for cases including massless particles will be simpler and we will not discuss it again.}
According to the definition of helicity amplitude~\cite{Jacob:1959at}, 
the polarization of the initial particle ($\lambda_1=\sigma_1$) is along the $z$-axis, and the polarizations of the final particles ($\lambda_2$ and $\lambda_3$) are along their respective motion directions.
One has
\begin{align}
	{\cal H}^{\lambda_2\lambda_3}_{\lambda_1}\left(\mathbf{k}_1,\mathbf{p}^*_2,\mathbf{p}^*_3;L,S\right) &=\, {\cal H}^{\lambda_2\lambda_3}_{\sigma_1}\left(\mathbf{k}_1,\mathbf{p}^*_2,\mathbf{p}^*_3;L,S\right)\notag\\
	&=\, D^{(s_2)\lambda_2}_{\sigma_2}\left(R_{\hat{\mathbf{p}}^*_2}\right)\,D^{(s_3)\lambda_3}_{\sigma_3}\left(R_{\hat{\mathbf{p}}^*_3}\right)\,{\cal A}_{\sigma_1}^{\sigma_2\sigma_3}\left(\mathbf{k}_1,\mathbf{p}^*_2,\mathbf{p}^*_3;L,S\right).
	\label{eq:helicity_amplitude_def}
\end{align}
From eq.~\eqref{eq:t_tilde}, the angular dependence of ${\cal A}$ comes from the relative momentum $\left(p_2^*-p_3^*\right)$, so one can separate the angular variables through a rotation transformation as follows,
\begin{align}
	\tilde{t}^{(L)}_{\alpha^L}(\mathbf{k}_1,\mathbf{p}^*_2-\mathbf{p}^*_3)&=\, P^{\mu_1\cdots\mu_L}_{\alpha^L}(\mathbf{k}_1;L)\,(p^*_2-p^*_3)_{\mu_1}\cdots (p^*_2-p^*_3)_{\mu_L}\notag\\
	&=\, P^{\mu_1\cdots\mu_L}_{\alpha^L}(\mathbf{k}_1;L)\,D_{\mu_1}^{~~\nu_1}\left(R_{\hat{\mathbf{p}}^*_2}\right)\,(\bar{p}^*_2-\bar{p}^*_3)_{\nu_1}\cdots D_{\mu_L}^{~~\nu_L}\left(R_{\hat{\mathbf{p}}^*_2}\right)\,(\bar{p}^*_2-\bar{p}^*_3)_{\nu_L}\notag\\
	&=\, D_{\alpha^L}^{~~\beta^L}\left(R_{\hat{\mathbf{p}}^*_2}\right)\,P^{\mu_1\cdots\mu_L}_{\beta^L}(\mathbf{k}_1;L)\,(\bar{p}^*_2-\bar{p}^*_3)_{\mu_1}\cdots(\bar{p}^*_2-\bar{p}^*_3)_{\mu_L}\notag\\
	&=\, D_{\alpha^L}^{~~\beta^L}\left(R_{\hat{\mathbf{p}}^*_2}\right)\,\tilde{t}^{(L)}_{\beta^L}(\mathbf{k}_1,\bar{\mathbf{p}}^*_2-\bar{\mathbf{p}}^*_3),
	\label{eq:t_tilde_separation}
\end{align}
where $(\bar{p}^*_2-\bar{p}^*_3)$ is the relative momentum along $z$-axis;
from the second line to the third line, the rotational invariance of spin projection tensor has been used.
Similarly, the rotation transformation matrix $D_{\alpha^L}^{~~\beta^L}\left(R_{\hat{\mathbf{p}}^*_2}\right)$ can be continuously transferred to the three polarization indices $\sigma_i~(i=1,2,3)$ and one obtains
\begin{align}
	{\cal A}_{\sigma_1}^{\sigma_2\sigma_3}(\mathbf{k}_1,\mathbf{p}^*_2,\mathbf{p}^*_3;L,S)= D^{(s_1)\sigma_1'}_{\sigma_1}\left(R_{\hat{\mathbf{p}}^*_2}\right)D^{(s_2)\sigma_2}_{\sigma_2'}\left(R^{-1}_{\hat{\mathbf{p}}^*_2}\right)D^{(s_3)\sigma_3}_{\sigma_3'}\left(R^{-1}_{\hat{\mathbf{p}}^*_2}\right)
	{\cal A}_{\sigma_1'}^{\sigma_2'\sigma_3'}(\mathbf{k}_1,\bar{\mathbf{p}}^*_2,\bar{\mathbf{p}}^*_3;L,S).
	\label{eq:amplitude_angle_dependence}
\end{align}
Then, substituting eq.~\eqref{eq:amplitude_angle_dependence} into eq.~\eqref{eq:helicity_amplitude_def}, one obtains
\begin{align}
	{\cal H}^{\lambda_2\lambda_3}_{\sigma_1}\left(\mathbf{k}_1,\mathbf{p}^*_2,\mathbf{p}^*_3;L,S\right)
	&=\,D^{(s_1)\sigma_1'}_{\sigma_1}\left(R_{\hat{\mathbf{p}}^*_2}\right)\,D^{(s_3)\lambda_3}_{\sigma_3}\left(R^{-1}_{\hat{\mathbf{p}}^*_2}\cdot R_{\hat{\mathbf{p}}^*_3}\right)\,{\cal A}_{\sigma_1'}^{\lambda_2\sigma_3}(\mathbf{k}_1,\bar{\mathbf{p}}^*_2,\bar{\mathbf{p}}^*_3;L,S)\notag\\
	&\equiv\,e^{i\Theta\left(R_{\hat{\mathbf{p}}^*_2},R_{\hat{\mathbf{p}}^*_3}\right)}\,D^{(s_1)\sigma_1'}_{\sigma_1}\left(R_{\hat{\mathbf{p}}^*_2}\right)\,{\cal F}_{\sigma_1'}^{\lambda_2\,\lambda_3}(\mathbf{k}_1,|\mathbf{p}^*_2|,|\mathbf{p}^*_3|;L,S),
	\label{eq:helicity_coupling_amplitude}
\end{align}
where $e^{i\Theta\left(R_{\hat{\mathbf{p}}^*_2},R_{\hat{\mathbf{p}}^*_3}\right)}$ is a global phase factor which depends on the phase convention; and
${\cal F}^{\lambda_2\lambda_3}_{\sigma_1}\left(\mathbf{k}_1,|\mathbf{p}^*_2|,|\mathbf{p}^*_3|;L,S\right)$ 
$=$ ${\cal A}_{\sigma_1}^{\lambda_2\,-\lambda_3}(\mathbf{k}_1,\bar{\mathbf{p}}^*_2,\bar{\mathbf{p}}^*_3;L,S)$ is called helicity-coupling amplitude~\cite{PhysRevD.48.1225}.
Eqs.~\eqref{eq:helicity_amplitude_def}, \eqref{eq:amplitude_angle_dependence} and \eqref{eq:helicity_coupling_amplitude} clearly show the relation between the helicity amplitude, the helicty coupling amplitude and the partial wave amplitude obtained from the covariant $L$-$S$ scheme.
%


\section{Examples of deriving Lorentz covariant partial wave amplitudes}
\label{sec:app}

In this section, we provide specific examples of two-body decay processes to further illustrate our scheme.


\subsection{Spin-one particle to two-body system of spin-one and spin-zero}
\label{sec:ex_1_10}

In this subsection, we employ eq.~\eqref{eq:general_LS_expand_1_motion} to obtain the Lorentz covariant partial wave amplitudes corresponding to the decay process $1(s_1=1)\to 2(s_2=1) + 3(s_3=0)$ (where all particles are massive).

Firstly, specific representations of $L_p$ must be selected to express the spin wave functions under consideration.
Here, recalling eq.~\eqref{eq:massive_particle_rep} and the conventions in table~\ref{tab:symbol_conventions}, one has 
\begin{align}
	\left[\alpha_1\right] = \left[\mu^1\right] &~~\Rightarrow~~ \bar{u}^{\alpha_1}_{\sigma_1}\left(\mathbf{k}_1;\left[\mu^1\right],1\right) = U^{\alpha_1}_{lr} \left(C^{\frac12\frac12}_1\right)^{lr}_{\sigma_1},\notag\\
	\left[\alpha_2\right] = \left[\mu^1\right] &~~\Rightarrow~~ u_{\alpha_2}^{\sigma_2}\left(\mathbf{k}_2;\left[\mu^1\right],1\right) = U_{\alpha_2}^{lr} \left(C_{\frac12\frac12}^1\right)_{lr}^{\sigma_1},
	\label{eq:ex_1_10_symbol_spin_orbit}\\
	\left[\alpha_3\right] = \left[\mu^0\right] &~~\Rightarrow~~ u_{\alpha_3}^{\sigma_3}\left(\mathbf{k}_3;\left[\mu^0\right],1\right) = U_{\alpha_3}^{00} \left(C_{00}^0\right)_{00}^{\sigma_1}.\notag
\end{align}
Then, to obtain the Lorentz covariant coupling structure  $\Gamma_{\alpha_1}^{\alpha_2\alpha_3}(\mathbf{k}_1,\mathbf{p}^*_2,\mathbf{p}^*_3;L,S)$ based on eq.~\eqref{eq:general_LS_expand_1_rest_Gamma}, one needs to calculate the three spin projection tensors $P^{\alpha^L\alpha^S}_{\alpha_1}(\mathbf{k}_1;s_1,L,S)$, $P^{\alpha_2\alpha_3}_{\alpha^S}(\mathbf{k}_1;S,s_2,s_3)$ and $P^{\beta_1\cdots\beta_L}_{\alpha^L}(\mathbf{k}_1;L)$.
Let us discuss them one by one.

The spin projection tensor $P^{\alpha^L\alpha^S}_{\alpha_1}(\mathbf{k}_1;s_1,L,S)$ has three indices, where $\alpha_1$ is introduced in eq.~\eqref{eq:ex_1_10_symbol_spin_orbit}.
The indices $\alpha^{S}$ and $\alpha^{L}$ carry the representations that depend on the quantum number of total spin $S$ and orbital angular momentum $L$, respectively. Applying the selection rule of angular momentum, one can get
\begin{equation}
	|s_2-s_3|~\leq~S~\leq~s_2+s_3~\Rightarrow~S=1,\qquad
	|s_1-S|~\leq~L~\leq~s_1+S~\Rightarrow~L=0,1,2.
	\label{eq:ex_triangular_relation_L_1_10}
\end{equation}
Thus, one has $\left[\alpha^S\right]=\left[\mu^1\right]$ and $\left[\alpha^L\right]=\left[\mu^L\right]$. According to eq.~\eqref{eq:spin_proj_ten_JLS}, we obtain the explicit form of  $P^{\alpha^L\alpha^S}_{\alpha_1}(\mathbf{k}_1;s_1,L,S)$ as follows,
\begin{equation}
	P^{\alpha^L\alpha^S}_{\alpha_1}(\mathbf{k}_1;1,L,1) ~=~ \left(C_{1}^{L1}\right)_{\sigma_1}^{\sigma_L\sigma_S}\,u^{\sigma_1}_{\alpha_1}\left(\mathbf{k}_1;\left[\mu^1\right],1\right)\,\bar{u}^{\alpha^L}_{\sigma_L}\left(\mathbf{k}_1;\left[\mu^L\right],L\right)\, \bar{u}^{\alpha^S}_{\sigma_S}\left(\mathbf{k}_1;\left[\mu^1\right],1\right),
	\label{eq:ex_1_spin_proj_ten_1LS}
\end{equation}
where we have choosen $C_{\chi_{_1}\chi_{_S}\chi_{_L}}=1$ for simplicity, since all $\chi_i~(i=1,S,L)$ are fixed.

The spin projection tensor $P_{\alpha^S}^{\alpha_2\alpha_3} \left(\mathbf{k}_1;S,s_2,s_3\right)$ has three indices.
From the above discussion, we have $[\alpha_1]=[\alpha_2]=\left[\alpha^S\right]=\left[\mu^1\right]$ and $[\alpha_3]=\left[\alpha^{L=0}\right]=\left[\mu^0\right]$, thus, the explicit form of $P_{\alpha^S}^{\alpha_2\alpha_3} \left(\mathbf{k}_1;1,1,0\right)$ is the same as $P^{\alpha^L\alpha^S}_{\alpha_1}(\mathbf{k}_1;1,1,L)$ with $L=0$.

The spin projection tensor $P^{\beta_1\cdots\beta_L}_{\alpha^L}(\mathbf{k}_1;L)$ has $L+1$ indices, which can be obtained recursively according to eq.~\eqref{eq:t_tilde}. The explicit forms are as follows,
\begin{align}
	L=0 ~&:\qquad~ P_{\alpha^0}(\mathbf{k}_1;0) ~=~ u_{\alpha^0}^{0}\left(\mathbf{k}_1;\left[\mu^0\right],0\right),\notag\\
	L=1 ~&:\qquad~ 	P^{\beta_1}_{\alpha^1}(\mathbf{k}_1;1)= P^{\beta_1\alpha^{0}}_{\alpha^1}(\mathbf{k}_1;1,1,0)~ P_{\alpha^{0}}(\mathbf{k}_1;0),\notag\\
	L=2 ~&:\qquad~ 	P^{\beta_1\beta_2}_{\alpha^2}(\mathbf{k}_1;2)= P^{\beta_1\alpha^{1}}_{\alpha^2}(\mathbf{k}_1;2,1,1)~ P^{\beta_2}_{\alpha^{1}}(\mathbf{k}_1;1),
	\label{eq:ex_1_orbit_wave_func}
\end{align}
where $P^{\beta_1\alpha^{0}}_{\alpha^1}(\mathbf{k}_1;1,1,0)$ is the same as $P_{\alpha^S}^{\alpha_2\alpha_3} \left(\mathbf{k}_1;1,1,0\right)$; and $P^{\beta_1\alpha^{1}}_{\alpha^2}(\mathbf{k}_1;2,1,1)$ is as follows,
\begin{equation}
	P^{\beta_1\alpha^{1}}_{\alpha^2}(\mathbf{k}_1;2,1,1)=\left(C_{2}^{11}\right)_{\sigma}^{\sigma'\sigma''}\,u^{\sigma}_{\alpha_2}\left(\mathbf{k}_1;\left[\mu^2\right],2\right)\, \bar{u}^{\beta_1}_{\sigma'}\left(\mathbf{k}_1;\left[\mu^1\right],1\right)\,\bar{u}^{\alpha^1}_{\sigma''}\left(\mathbf{k}_1;\left[\mu^1\right],1\right).
\end{equation}

Combining the aforementioned spin projection tensors, one can obtain Lorentz covariant coupling structures with different orbital angular momentum $L$, expressed as follows,
\begin{equation}
	\Gamma_{\alpha_1}^{\alpha_2\alpha_3}(\mathbf{k}_1,\mathbf{p}^*_2,\mathbf{p}^*_3;L,1)=\,P^{\alpha^L\alpha^S}_{\alpha_1}(\mathbf{k}_1;1,L,1)\,P^{\alpha_2\alpha_3}_{\alpha^S}(\mathbf{k}_1;1,1,0)\,\tilde{t}^{(L)}_{\alpha^L}(\mathbf{k}_1,\mathbf{p}^*_2-\mathbf{p}^*_3).
\end{equation}
Further, one can use the similarity transformations as shown in eq.~\eqref{eq:appd_spacetime_chiral} to get these coupling structures in space-time representation as follows,
\begin{equation}
	\Gamma_{\mu}^{\nu}(\mathbf{k}_1,\mathbf{p}^*_2,\mathbf{p}^*_3;L,1)~=~U_{\mu}^{~~\alpha_1}~\left(U^{-1}\right)^{~~\nu}_{\alpha_2}~\Gamma_{\alpha_1}^{\alpha_2\alpha_3}(\mathbf{k}_1,\mathbf{p}^*_2,\mathbf{p}^*_3;L,1),
\end{equation}
where we have droped the index $\alpha_3$, since $[\alpha_3]=\left[\mu^0\right]$ is the identity representation.
The explicit forms can be obtained as follows,
\begin{align}
	\Gamma_{\mu}^{\nu}(\mathbf{k}_1,\mathbf{p}^*_2,\mathbf{p}^*_3;0,1)~&=~ -g_{\mu}^{~~\nu}~+~\frac{k_{1\mu}k_1^{\nu}}{m_1^2},\notag\\
	\Gamma_{\mu}^{\nu}(\mathbf{k}_1,\mathbf{p}^*_2,\mathbf{p}^*_3;1,1)~&=~ i\,g_{\mu\mu'}\,\left(\frac{k_1}{m_1}\right)_{\nu'}~U_{\rho'}^{~~\alpha^1}~ \tilde{t}^{(1)}_{\alpha^1}(\mathbf{k}_1,\mathbf{p}^*_2-\mathbf{p}^*_3)\,\epsilon^{\mu'\nu'\rho'\nu}\notag\\
	\Gamma_{\mu}^{\nu}(\mathbf{k}_1,\mathbf{p}^*_2,\mathbf{p}^*_3;2,1)~&=~U_{\mu}^{~~\alpha_1}~\left(U^{-1}\right)^{~~\nu}_{\alpha_2}~T_{\alpha_1}^{\alpha_2\alpha^2}~\tilde{t}^{(2)}_{\alpha^2}(\mathbf{k}_1,\mathbf{p}^*_2-\mathbf{p}^*_3),
	\label{eq:ex_1_Gamma_L_0}
\end{align}
where $k_{1\mu}=(m_1,0,0,0)$; $T_{\alpha_1}^{\alpha_2\alpha^2}$ is an order-3 IRT of $L_p$ which can be found in appendix~\ref{appd:ex_irten_Lp}.

Finally, by combining eqs.~\eqref{eq:general_LS_expand_1_motion} and \eqref{eq:ex_1_Gamma_L_0}, one can obtain the Lorentz covariant partial wave amplitudes as follows,
\begin{equation}
	\mathcal{A}_{\sigma_1}^{\sigma_2}(\mathbf{p}_1,\mathbf{p}_2,\mathbf{p}_3;L,1) ~=~ \bar{\epsilon}_{\sigma_1}^{\mu}\left(\mathbf{p}_1\right)\,\Gamma_{\mu}^{\rho}(\mathbf{p}_1,\mathbf{p}_2,\mathbf{p}_3;L,1)\,D^{~~\nu}_{\rho}\left(h_{\mathbf{p}_1}\cdot h_{\mathbf{p}_2^*}^{-1}\cdot h_{\mathbf{p}_1}^{-1}\right)\,\epsilon^{\sigma_2}_{\nu}\left(\mathbf{p}_2\right),
	\label{eq:ex_1_amplitude}
\end{equation}
where $\epsilon_\mu^{\sigma}(\mathbf{p})=U_\mu^{~~\alpha} u_\alpha^\sigma\left(\mathbf{p};\left[\mu^1\right];1\right)$ and $\bar{\epsilon}^\mu_{\sigma}(\mathbf{p})=\left(U^{-1}\right)^{~~\mu}_{\alpha} \bar{u}^\alpha_\sigma\left(\mathbf{p};\left[\mu^1\right];1\right)$ are covariant and contravariant polarization vectors, respectively; $D^{~~\nu}_{\rho}\left(h_{\mathbf{p}_1}\cdot h_{\mathbf{p}_2^*}^{-1}\cdot h_{\mathbf{p}_1}^{-1}\right)$ is a Lorentz transformation matrix.
If one removes $D^{~~\nu}_{\rho}\left(h_{\mathbf{p}_1}\cdot h_{\mathbf{p}_2^*}^{-1}\cdot h_{\mathbf{p}_1}^{-1}\right)$ in eq.~\eqref{eq:ex_1_amplitude}, the partial wave amplitude will be exactly the same as the result of ref.~\cite{Zou:2002ar}.
In fact, as shown in eq.~\eqref{eq:general_LS_expand_1_motion}, if $D_\alpha^{~~\beta}\left(R_{1i}\right)$ is replaced by $D_\alpha^{~~\beta}\left(h_{\mathbf{p}_{i}^*}\cdot R_{1i}\right)~(i=1,2)$, such partial wave amplitudes for fixed $L$ and $S$ are exactly the same as that in ref.~\cite{Zou:2002ar}. Now they are only consistent at the two-body threshold of final state, i.e., $|\mathbf{p}^*_{2}|/E^*_{2}\to0$ and then $D^{~~\nu}_{\rho}\left(h_{\mathbf{p}_1}\cdot h_{\mathbf{p}_2^*}^{-1}\cdot h_{\mathbf{p}_1}^{-1}\right)\to \delta^{~~\nu}_{\mu}$.
However, when both methods include all partial wave amplitudes for $L$-$S$ quantum numbers, they are actually equivalent because they both include all independent Lorentz covariant structures and just use different complete bases to parameterize the amplitudes. In other words, the definition of $L$-$S$ quantum number is different in two methods, while our definition as shown in eq.~\eqref{eq:general_LS_expand_1_motion} is exactly the same as them in ref.~\cite{Jacob:1959at}.
%


\subsection{Spin-half particle to two-body system of spin-half and spin-one}
\label{sec:ex_12_121}

In this subsection, we employ eq.~\eqref{eq:general_LS_expand_1_motion} to obtain the Lorentz covariant partial wave amplitudes corresponding to the decay process $1(s_1=\frac12)\to 2(s_2=\frac12)+3(s_3=1)$ (where all particles are massive).

Firstly, specific representations of $L_p$ must be selected to express the spin wave functions under consideration.
Recalling eq.~\eqref{eq:massive_particle_rep} and the conventions in table~\ref{tab:symbol_conventions}, one has
\begin{align}
	\left[\alpha_1\right] = \left[a^1\right] &~~\Rightarrow~~ \bar{u}^{\alpha_1}_{\sigma_1}\left(\mathbf{k}_1;\chi,\frac12\right) =\left\{ \begin{array}{ll}
		~\left(U_R\right)^{\alpha_1}_{lr} \left(C^{0\frac12}_{\frac12}\right)^{lr}_{\sigma_1}\quad\text{for}~\chi=\left[a^1_R\right]\\
		\\
		~\left(U_L\right)^{\alpha_1}_{lr} \left(C^{\frac12 0}_{\frac12}\right)^{lr}_{\sigma_1}\quad\text{for}~\chi=\left[a^1_L\right]
	\end{array}\right.,  \notag\\
	\left[\alpha_2\right] = \left[a^1\right] &~~\Rightarrow~~ u_{\alpha_2}^{\sigma_2}\left(\mathbf{k}_2;\chi,\frac12\right) =\left\{ \begin{array}{ll}
		~\left(U_L\right)_{\alpha_2}^{lr} \left(C_{\frac12 0}^{\frac12}\right)_{lr}^{\sigma_2}\quad\text{for}~\chi=\left[a^1_L\right]\\
		\\
		~\left(U_R\right)_{\alpha_2}^{lr} \left(C_{\frac12 0}^{\frac12}\right)_{lr}^{\sigma_2}\quad\text{for}~\chi=\left[a^1_R\right]
	\end{array}\right.,
	\label{eq:ex_12_121_symbol_spin_orbit}\\
	\left[\alpha_3\right] = \left[\mu^1\right] &~~\Rightarrow~~ u_{\alpha_3}^{\sigma_3}\left(\mathbf{k}_3;\left[\mu^1\right],1\right) = U_{\alpha_3}^{lr} \left(C_{\frac12\frac12}^1\right)_{lr}^{\sigma_3}.\notag
\end{align} 
Then, to obtain the Lorentz covariant coupling structure $\Gamma_{\alpha_1}^{\alpha_2\alpha_3}(\mathbf{k}_1,\mathbf{p}^*_2,\mathbf{p}^*_3;L,S)$ based on eq.~\eqref{eq:general_LS_expand_1_rest_Gamma}, one needs to calculate the three spin projection tensors $P^{\alpha^L\alpha^S}_{\alpha_1}(\mathbf{k}_1;s_1,L,S)$, $P^{\alpha_2\alpha_3}_{\alpha^S}(\mathbf{k}_1;S,s_2,s_3)$ and $P^{\beta_1\cdots\beta_L}_{\alpha^L}(\mathbf{k}_1;L)$.
Let us discuss them one by one.

The spin projection tensor $P^{\alpha^L\alpha^S}_{\alpha_1}(\mathbf{k}_1;s_1,L,S)$ has three indices, where $\alpha_1$ is introduced in eq.~\eqref{eq:ex_12_121_symbol_spin_orbit}.
Applying the selection rule of angular momentum, one can get
\begin{equation}
	|s_2-s_3|~\leq~S~\leq~s_2+s_3~\Rightarrow~S=\frac12,\frac32,\qquad
	|s_1-S|~\leq~L~\leq~s_1+S~\Rightarrow~L=0,1,2.
	\label{eq:ex_triangular_relation_L_12_121}
\end{equation}
Thus, one has $\left[\alpha^S\right]=\left[a^{2S}\right]$ and $\left[\alpha^L\right]=\left[\mu^L\right]$. According to eq.~\eqref{eq:spin_proj_ten_JLS}, we obtain the explicit form of  $P^{\alpha^L\alpha^S}_{\alpha_1}(\mathbf{k}_1;s_1,L,S)$ as follows,
\begin{align}
	P^{\alpha^L\alpha^S}_{\alpha_1}\left(\mathbf{k}_1;\frac12,L,S\right)=\sum_{\chi_{_1}=\left[a^1_{L/R}\right],\chi_{_S}=\left[a^{2S}_{L/R}\right]}~&C_{\chi_{_1}\chi_{_S}}~ \left(C_{\frac12}^{LS}\right)_{\sigma_1}^{\sigma_L\sigma_S}~u^{\sigma_1}_{\alpha_1}\left(\mathbf{k}_1;\chi_{_1},\frac12\right)\notag\\
	&\times\bar{u}^{\alpha^L}_{\sigma_L}\left(\mathbf{k}_1;\left[\mu^L\right],L\right)~ \bar{u}^{\alpha^S}_{\sigma_S}\left(\mathbf{k}_1;\chi_{_S},S\right),
	\label{eq:ex_2_spin_proj_ten_1LS}
\end{align}
where we have droped $\chi_{_L}$ since it is fixed; $C_{\chi_{_1}\chi_{_S}}$s are some indeterminate coefficients as mentioned in eq.~\eqref{eq:spin_proj_ten_JLS}, which represents the difference of coupling strength between left and right-handed spinors.
If coupling structure with conserved quantities such as $\mathcal{P}$-parity and $\mathcal{C}$-parity is required, these $C_{\chi_{_1}\chi_{_S}}$s will be limited by additional conditions.

The spin projection tensor $P_{\alpha^S}^{\alpha_2\alpha_3} \left(\mathbf{k}_1;S,s_2,s_3\right)$ has three indices, where $[\alpha_2]=\left[a^1\right]$, $[\alpha_3]=\left[\mu^1\right]$ and $\left[\alpha^S\right]=\left[a^{2S}\right]$, which are chosen from the $P^{\alpha^L\alpha^S}_{\alpha_1}(\mathbf{k}_1;s_1,L,S)$ above. Again, according to eq.~\eqref{eq:spin_proj_ten_JLS}, we obtain the explicit form of $P_{\alpha^S}^{\alpha_2\alpha_3} \left(\mathbf{k}_1;S,s_2,s_3\right)$ as follows,
\begin{align}
	P_{\alpha^S}^{\alpha_2\alpha_3} \left(\mathbf{k}_1;S,\frac12,1\right)=\sum_{\chi_{_S}=\left[a^{2S}_{L/R}\right],\chi_{_2}=\left[a^1_{L/R}\right]}~&C_{\chi_{_S}\chi_{_2}}~ \left(C_{S}^{\frac12 1}\right)_{\sigma_S}^{\sigma_2\sigma_3}~u^{\sigma_S}_{\alpha^S}\left(\mathbf{k}_1;\chi_{_S},S\right)\notag\\
	&\times\bar{u}^{\alpha_2}_{\sigma_2}\left(\mathbf{k}_1;\chi_{_2},\frac12\right)~ \bar{u}^{\alpha_3}_{\sigma_3}\left(\mathbf{k}_1;\left[\mu^1\right],1\right),
\end{align}
where we have droped $\chi_{_3}$ since it is fixed; $C_{\chi_{_S}\chi_{_2}}$s are indeterminate coefficients similar to $C_{\chi_{_1}\chi_{_S}}$s in eq.~\eqref{eq:ex_2_spin_proj_ten_1LS}.

The spin projection tensor $P^{\beta_1\cdots\beta_L}_{\alpha^L}(\mathbf{k}_1;L)$ only depends on angular momentum $L$.
Eq.~\eqref{eq:ex_triangular_relation_L_12_121} gives the possible values of $L$, which is the same as that in eq.~\eqref{eq:ex_triangular_relation_L_1_10}. 
Thus, $P^{\mu_1\cdots\mu_L}_{\mu^L}(\mathbf{k}_1;L)$ here is the same as that in eq.~\eqref{eq:ex_1_orbit_wave_func}.

Then, Lorentz covariant coupling structures with different $(L,S)$ combinations are obtained by combining the above three spin projection tensors together as follows,
\begin{equation}
	\Gamma_{\alpha_1}^{\alpha_2\alpha_3}(\mathbf{k}_1,\mathbf{p}^*_2,\mathbf{p}^*_3;L,S)=\,P^{\alpha^L\alpha^S}_{\alpha_1}\left(\mathbf{k}_1;\frac12,L,S\right)\,P_{\alpha^S}^{\alpha_2\alpha_3} \left(\mathbf{k}_1;S,\frac12,1\right)\,\tilde{t}^{(L)}_{\alpha^L}(\mathbf{k}_1,\mathbf{p}^*_2-\mathbf{p}^*_3).
	\label{eq:ex_2_Gamma_SL_all}
\end{equation}
Finally, by combining eqs.~\eqref{eq:general_LS_expand_1_motion} and \eqref{eq:ex_2_Gamma_SL_all}, one can get the explicit form of these partial wave amplitudes as follows,
\begin{align}
	\mathcal{A}_{\sigma_1}^{\sigma_2\sigma_3}(\mathbf{p}_1,\mathbf{p}_2,\mathbf{p}_3;L,S) ~=~& \bar{u}_{\sigma_1}^{\alpha_1}\left(\mathbf{k}_1;\left[a^1\right],\frac12^{P_1}\right)\,\Gamma_{\alpha_1}^{\alpha_2'\alpha_3'}(\mathbf{k}_1,\mathbf{p}^*_2,\mathbf{p}^*_3;L,S)\, D^{~~\alpha_2}_{\alpha_2'}\left(R_{12}\right)\notag\\
	&\times D^{~~\alpha_3}_{\alpha_3'}\left(R_{13}\right)\,u^{\sigma_2}_{\alpha_2}\left(\mathbf{k}_2;\left[a^1\right],\frac12^{P_2}\right)\,u^{\sigma_3}_{\alpha_3}\left(\mathbf{k}_3;\left[\mu^1\right];1\right),
	\label{eq:ex_2_amplitude}
\end{align}
where $u^{\sigma}_{\alpha}\left(\mathbf{k}_i;[\alpha],s^P\right)/\bar{u}_{\sigma}^{\alpha}\left(\mathbf{k}_i;[\alpha],s^P\right)~(i=1,2)$ is covariant/contravariant spin wave function for particle with definite parity $P$ as defined in eq.~\eqref{eq:spin_wave_func_parity_eigenstate}.

In addition, a Lorentz covariant amplitude are expressed as contraction of tensors containing only Dirac spinor and Lorentz four-vector indices in the conventional form. 
However, as we have seen in subsections~\ref{sec:ex_1_10} and \ref{sec:ex_12_121}, the conventional form, which will become more and more complicated with the increase of spin, is unnecessary.
Therefore, detailed results of the amplitudes (eq.~\eqref{eq:ex_2_amplitude}) in the conventional form are not presented. Instead, a brief discussion of the relation between the different forms of spin wave function is included in appendix~\ref{appd_relation_spin_wave_func}.


\subsection{Three-particle partial wave amplitude including massless particles}
\label{sec:massless_LS_amp}

In this subsection, we reconsider the example in subsection~\ref{sec:ex_1_10} with particle-2 is massless (such as photon) and make a general discussion on three-particle partial wave amplitude including massless particles. 
According to eq.~\eqref{eq:self_conj_rep_spin_helicity}, we take $[\alpha_2]=(1,0)\oplus(0,1)\equiv[\zeta]$, and use the same representations for particle-1 and particle-3 as in eq.~\eqref{eq:ex_1_10_symbol_spin_orbit}.
From eq.~\eqref{eq:spin_wave_func_massless_explicit}, the left and right-handed helicity wave functions of photon can be written as follows,
\begin{align}
	u_{\zeta}^{\sigma_2}\left(\mathbf{k}_2;(1,0),1\right)=\delta_{\zeta}^{1}\,\delta_{-1}^{\sigma_2},\qquad
	u_{\zeta}^{\sigma_2}\left(\mathbf{k}_2;(0,1),1\right)=\delta_{\zeta}^{6}\,\delta_{1}^{\sigma_2}.
	\label{eq:helicity_proj_ten_ex_1}
\end{align}
It is worth emphasizing that $\sigma_2~(=\pm1)$ in the above equation refers to helicity rather than polarization components along a fixed direction.
According to eq.~\eqref{eq:spin_wave_func_parity_eigenstate},
the above two helicity wave functions can be also written as the following form of parity-eigenstate,
\begin{align}
	\text{Positive parity : }~ u_{\zeta}^{\sigma_2}\left(\mathbf{k}_2;[\zeta],1^+\right) ~&=~
	u_{\zeta}^{\sigma_2}\left(\mathbf{k}_2;(1,0),1\right) +u_{\zeta}^{\sigma_2}\left(\mathbf{k}_2;(0,1),1\right),\notag\\
	\text{Negative parity : }~ u_{\zeta}^{\sigma_2}\left(\mathbf{k}_2;[\zeta],1^-\right) ~&=~
	u_{\zeta}^{\sigma_2}\left(\mathbf{k}_2;(1,0),1\right) -u_{\zeta}^{\sigma_2}\left(\mathbf{k}_2;(0,1),1\right).
	\label{eq:spin_wave_func_massless_parity}
\end{align}
To obtain the Lorentz covariant coupling structure  $\Gamma_{\alpha_1}^{\zeta\alpha_3}(\mathbf{k}_1,\mathbf{p}^*_2,\mathbf{p}^*_3;L,S)$ based on eq.~\eqref{eq:general_LS_expand_1_rest_Gamma}, one needs to calculate three spin projection tensors $P^{\alpha^L\alpha^S}_{\alpha_1}(\mathbf{k}_1;s_1,L,S)$, $P^{\zeta\alpha_3}_{\alpha^S}(\mathbf{k}_1;S,s_2,s_3)$ and $P^{\beta_1\cdots\beta_L}_{\alpha^L}(\mathbf{k}_1;L)$.
Let us discuss them one by one.

The specific form of $P^{\alpha^L\alpha^S}_{\alpha_1}(\mathbf{k}_1;s_1,L,S)$ and $P^{\beta_1\cdots\beta_L}_{\alpha^L}(\mathbf{k}_1;L)$ solely depends on the spin-orbital quantum numbers $L$ and $S$, which are identical to those presented in eqs.~\eqref{eq:ex_1_spin_proj_ten_1LS} and \eqref{eq:ex_1_orbit_wave_func}.
Using eq.~\eqref{eq:spin_proj_ten_JLS}, we can derive the explicit form of $P^{\zeta\alpha_3}_{\alpha^S}(\mathbf{k}_1;S,s_2,s_3)$ as follows,
\begin{equation}
	P^{\zeta\alpha^3}_{\alpha_S}(\mathbf{k}_1;1,1,0) =\sum_{\chi=(1,0),(0,1)}C_{\chi} \left(C_{1}^{10}\right)_{\sigma_S}^{\sigma\sigma_3}\,u^{\sigma_S}_{\alpha^S}\left(\mathbf{k}_1;\left[\mu^1\right],1\right)\,\bar{u}^{\zeta}_{\sigma}\left(\mathbf{k}_1;\chi,1\right)\, \bar{u}^{\alpha_3}_{\sigma_3}\left(\mathbf{k}_1;\left[\mu^0\right],0\right),
	\label{eq:ex_3_spin_proj_ten_1LS}
\end{equation}
where we have droped $\chi_i~(i=S,3)$ since they are fixed; $C_{\chi}$s are indeterminate coefficients that are similar to $C_{\chi_{_1}\chi_{_S}}$s in eq.~\eqref{eq:ex_2_spin_proj_ten_1LS};
$\bar{u}^{\zeta}_{\sigma}\left(\mathbf{k}_1;\chi,1\right)$ denotes the contravariant spin wave function for a massive particle in the self-conjugate REREP $[1,0]$, which can be obtained from eq.~\eqref{eq:self_conj_rep_spin_wave_func}.

One can obtain Lorentz covariant coupling structures with different orbital angular momentum $L$ by combining the aforementioned spin projection tensors. It can be expressed as follows,
\begin{equation}
	\Gamma_{\alpha_1}^{\zeta}(\mathbf{k}_1,\mathbf{p}^*_2,\mathbf{p}^*_3;L,1)=\,P^{\alpha^L\alpha^S}_{\alpha_1}(\mathbf{k}_1;1,L,1)\,P^{\zeta}_{\alpha^S}(\mathbf{k}_1;1,1,0)\,\tilde{t}^{(L)}_{\alpha^L}(\mathbf{k}_1,\mathbf{p}^*_2-\mathbf{p}^*_3),
	\label{eq:ex_3_Gamma_L}
\end{equation}
where we have droped $\alpha_3$ for simplicity, since particle-3 is a scalar.

Finally, by combining eqs.~\eqref{eq:general_LS_expand_1_motion} and \eqref{eq:ex_3_Gamma_L}, we obtain the Lorentz covariant partial wave amplitudes satisfying gauge invariance as follows,
\begin{equation}
	(\mathcal{A}^{\pm})_{\sigma_1}^{\sigma_2}(\mathbf{p}_1,\mathbf{p}_2,\mathbf{p}_3;L,1) = \bar{u}_{\sigma_1}^{\alpha_1}\left(\mathbf{k}_1;\left[\mu^1\right],1\right)\,\Gamma_{\alpha_1}^{\zeta'}(\mathbf{k}_1,\mathbf{p}^*_2,\mathbf{p}^*_3;L,1)\,D^{~~\zeta}_{\zeta'}\left(\tilde{R}_{12}\right)\,u^{\sigma_2}_{\zeta}\left(\mathbf{k}_2;[\zeta];1^\pm\right).
	\label{eq:ex_3_amplitude}
\end{equation}

Furthermore, according to eq.~\eqref{eq:helicity_coupling_amplitude}, the helicity-coupling amplitudes which are independent on angular variables can be extracted from partial wave amplitudes in eq.~\eqref{eq:ex_3_amplitude} as follows,
\begin{align}
	{({\cal F}^\pm)}_{\sigma_1}^{\lambda_2}(\mathbf{k}_1,|\mathbf{p}^*_2|,|\mathbf{p}^*_3|;0,1) &\propto \begin{pmatrix}
		~\,1\,~ & 0 & 0 \\
		0 & 0 & 0 \\
		0 & 0 & \pm1\,~
	\end{pmatrix}_{\sigma_1}^{~~\lambda_2},\notag\\
	{({\cal F}^\pm)}_{\sigma_1}^{\lambda_2}(\mathbf{k}_1,|\mathbf{p}^*_2|,|\mathbf{p}^*_3|;1,1) &\propto \begin{pmatrix}
		\,~1~\, & 0 & 0 \\
		0 & 0 & 0 \\
		0 & 0 & \mp1~\,
	\end{pmatrix}_{\sigma_1}^{~~\lambda_2},\notag\\
	{({\cal F}^\pm)}_{\sigma_1}^{\lambda_2}(\mathbf{k}_1,|\mathbf{p}^*_2|,|\mathbf{p}^*_3|;2,1) &\propto \begin{pmatrix}
		\,~1~\, & 0 & 0 \\
		0 & 0 & 0 \\
		0 & 0 & \pm1~\,
	\end{pmatrix}_{\sigma_1}^{~~\lambda_2}.
\end{align}

It can be seen that only two of the above three amplitudes are linearly independent.
This example shows that, for process involving massless particles such as photons, with definite $L$-$S$ quantum numbers, we are able to write down the partial wave amplitudes which satisfy both Lorentz covariance and gauge invariance.
However, in general, these amplitudes are not all linearly independent (e.g., see the treatment of radiation decay of $\psi$ in ref.~\cite{Zou:2002ar}).
Therefore, to avoid the introduction of redundant fitting parameters in the PWA of experimental data, it is necessary to select the linearly independent terms from the partial wave amplitudes of all possible $L$-$S$ combinations.
To achieve this, one must determine the number of linearly independent terms in these partial wave amplitudes and select a complete basis from these amplitudes. 
We define vectors $V_I(s_1,s_2,s_3,L,S)$ based on the helicity-coupling amplitude definition given in Eq.~\eqref{eq:helicity_coupling_amplitude} as follows,
\begin{align}
	V_I(s_1,s_2,s_3,L,S)~&\equiv~V_{\,\sigma_1}^{\sigma_2\sigma_3}(s_1,s_2,s_3,L,S)~=~\left(C^{s_2s_3}_{S}\right)^{\sigma_2\sigma_3}_{\sigma_1}~\left(C^{SL}_{s_1}\right)^{\sigma_1\,0}_{\sigma_1},\notag\\
	\mathrm{dim}\,[I] ~&=~ \mathrm{dim}\,[\sigma_1]\times\mathrm{dim}\,[\sigma_2]\times\mathrm{dim}\,[\sigma_3],
\end{align}
where we have denoted the three spin polarization indices $\sigma_i~(i=1,2,3)$ as an index $I$.
The number of linearly independent terms in partial wave amplitudes equals the number of linearly independent vectors $V_I(s_1,s_2,s_3,L,S)$ of all possible $L$-$S$ combinations. 

Table~\ref{tab:massless_InDenNum_PWA} provides formulas for calculating the number of linearly independent partial wave amplitudes. To obtain a set of linearly independent partial wave amplitudes, one needs to select $L$-$S$ combinations of a given number (the number of linearly independent terms) from all possible $L$-$S$ combinations and ensure that the corresponding vectors $V_I(s_1,s_2,s_3,L,S)$ are linearly independent with each other. However, there is no unique selection criterion except that the number of $L$-$S$ combinations with even and odd $L$ must be equal.
To facilitate the selection process, we construct a weight function $W(s_1,s_2,s_3,L,S)$ (see appendix~\ref{appd:weight_func}) and suggest\footnote{This is only a suggestion, because any other $L$-$S$ combinations can also be selected, as long as these amplitudes are linearly independent.} that $L$-$S$ combinations with larger weight can be considered first when selecting $L$-$S$ combinations within requirements.

\begin{table}
	\centering
	\vspace{1mm}
	
	\begin{tabular}{c|c|c|c|c}
		\hline
		~~Range~~ & ~~$N_0(s_1;s_2,s_3)$~~ & ~~$N_1(s_1;s_2,s_3)$~~ & ~~$N_2(s_1;s_2,s_3)$~~ & ~~$N_3(s_1;s_2,s_3)$~~\\* 
		\hline
		(a) & \multirow{2}{*}{$(2s_1+1)(2s_3+1)$} & 0 & 0 & 0  \\* 
		\cline{1-1}\cline{3-5}
		(b) &                    & \multirow{2}{*}{$2(s_1-s_2+s_3+1)$} & \multirow{3}{*}{2} & 2  \\* 
		\cline{1-2}\cline{5-5}
		(c) & $n(s_1;s_2,s_3)$                  &                    &                    & 0  \\* 
		\cline{1-3}\cline{5-5}
		(d) & \multirow{2}{*}{$(2s_1+1)(2s_2+1)$} & \multirow{2}{*}{$2(2s_1+1)$} &                    & 2  \\* 
		\cline{1-1}\cline{4-5}
		(e) &                    &                    & 0                  & 0  \\* 
		\hline
		(f) & \multirow{2}{*}{$(2s_2+1)(2s_3+1)$} & \multirow{2}{*}{$2(2s_3+1)$} & \multirow{2}{*}{4} & 2  \\* 
		\cline{1-1}\cline{5-5}
		(g) &                    &                    &                    & 0  \\
		\hline
	\end{tabular}
	\caption{The number of linearly independent terms of three-particle amplitudes $(s_1\to s_2+s_3)$ is determined in seven different ranges: (a) $s_1<s_2-s_3$; (b) $s_1=s_2-s_3$; (c) $|s_2-s_3|<s_1<s_2+s_3$; (d) $s_1=s_3-s_2$; (e) $s_1<s_3-s_2$; (f) $s_1=s_2+s_3$; (g) $s_1>s_2+s_3$. $N_i(s_1;s_2,s_3)~(i=0,1,2,3)$ represents the number of linearly independent terms for four cases: ($i=0$) all particles are massive; ($i=1$) particle-2 is massless and $s_2\neq 0$; ($i=2$) both particle-2 and 3 are massless and $s_{2,3}\neq0$; ($i=3$) all particles are massless and $s_{1,2,3}\neq0$. The value of $n(s_1;s_2,s_3)$ is calculated as follows: $n(s_1;s_2,s_3)=-(s_1^2+s_2^2+s_3^2)+2(s_1s_2+s_2s_3+s_1s_3)+s_1+s_2+s_3+1$.}
	\label{tab:massless_InDenNum_PWA}
\end{table}


\section{Summary}
\label{sec:summary}

In this work, we generalized the covariant $L$-$S$ coupling scheme based on the IRTs of $L_p$ and its little groups (SO(3) and ISO(2)) which gives a general procedure for constructing the partial wave amplitude with obvious Lorentz covariant form.
The scheme is applicable to both massive and massless particles with arbitrary spins and to processes with or without the conservation of ${\cal P}$-parity, ${\cal C}$-parity, and so on. In addition, the partial wave amplitudes including gauge bosons proposed here automatically satisfy gauge invariance. Therefore, this scheme is useful for the PWA of strong and electroweak interaction processes.

We provided detailed derivations for calculating the partial wave amplitudes of three examples, including bosons, fermions, and photon. Through the example in subsection~\ref{sec:ex_1_10}, we show that a pure $L$-$S$ component in the corresponding partial wave amplitude labeled by $(L,S)$ is well-defined for any initial mass $m_1$, while in refs.~\cite{Zou:2002ar,Zou:2002yy,Dulat:2011rn}, this was only the case for threshold mass $m_1=m_2+m_3$ (non-relativistic limit). This point should be paid special attention when cross-checking the results of different PWA schemes.
Through the example in subsection~\ref{sec:ex_12_121}, we show that the conventional form for expressing Lorentz covariant amplitude is unnecessary. If needed, the IRTs of $L_p$ can be used to convert different forms of spin wave functions.
In subsection~\ref{sec:massless_LS_amp}, we show that the number of linearly independent partial wave amplitudes for cases that include massless particles is always less than the number of all possible $L$-$S$ combinations because of gauge invariance. To prevent the introduction of redundant fitting parameters, we introduce a weight function $W(s_1,s_2,s_3,L,S)$ to select a set of linearly independent complete bases from these partial wave amplitudes.

\newpage

\begin{acknowledgments}
	
	We thank useful discussions with Xiao-Yu Li, Xiang-Kun Dong and Feng-Kun Guo.
	This work is partly supported by the China Postdoctoral Science Foundation under Grants No. 119103S408 (H.J.J.),
	and by National Natural Science Foundation of China under Grants No.12175239, 12221005 (J.J.W.),
	and by the National Key R$\&$D Program of China under Contract No.
	2020YFA0406400 (J.J.W.),
	and by the NSFC under Grant No. 12070131001 (CRC110 cofunded by the DFG and NSFC), Grant
	No. 11835015, No. 12047503, and by the Chinese Academy of Sciences (CAS) under Grant
	No. XDB34030000 (B.S.Z.).
	
\end{acknowledgments}


\appendix

\section{Covariant tensor and irreducible tensor}
\label{appd:irrep_and_irten}

In this appendix, we give a brief introduction to COVT and IRT. 
A tensor $T_{b_1b_2\cdots}^{a_1a_2\cdots}$ is called a COVT of a group $G$, where the lower/upper indices correspond to covariant/contravariant components, if it transforms as follows,
$$	T_{b_1b_2\cdots}^{a_1a_2\cdots} ~\xrightarrow{g\in G}~ \tilde{T}_{b_1b_2\cdots}^{a_1a_2\cdots} ~=~ D_{b_1}^{~~b'_1}(g)~D_{b_2}^{~~b'_2}(g)~D^{a_1}_{~~a'_1}(g)~D^{a_2}_{~~a'_2}(g)~\cdots~ T_{b'_1b'_2\cdots}^{a'_1a'_2\cdots},$$
where $D_{a}^{~~b}(g)$ and $D^{a}_{~~b}(g)$ are the corresponding IRREP matrices of arbitrary group element $g(\in G)$.

There is a special class of COVTs which is invariant under group transformation, called invariant tensor (INVT). 
For a given group $G$, one can obtain an INVT $T_{b_1b_2\cdots}^{a_1a_2\cdots}$ by solving the following equation except for an overall factor,
\begin{equation}
	T_{b_1b_2\cdots}^{a_1a_2\cdots} ~\xrightarrow{g\in G}~ \tilde{T}_{b_1b_2\cdots}^{a_1a_2\cdots} ~=~ D_{b_1}^{~~b'_1}(g)~D_{b_2}^{~~b'_2}(g)~D^{a_1}_{~~a'_1}(g)~D^{a_2}_{~~a'_2}(g)~\cdots~ T_{b'_1b'_2\cdots}^{a'_1a'_2\cdots} ~=~ T_{b_1b_2\cdots}^{a_1a_2\cdots}~.
	\label{eq:def_of_irten}
\end{equation}  
Tensors can be divided by the number of indices, as order-$0$, order-$1$, order-$2$ and so on. 
Let us discuss INVTs order by order. 
Without loss of generality, we will only discuss INVTs with covariant indices.

\textbf{Order-0 INVT}: an order-0 INVT $T$ has no index and must be a scalar, which is invariant automatically under group transformation. 
This is a trivial case. 

\textbf{Order-1 INVT}: an order-1 INVT $T_a$ is a vector, which is a set of bases that carry the IRREP $D_{a}^{~~b}(G)$, while it is also invariant under group transformation. 
The IRREP must be the identity representation which returns to the case of order-0 INVT. Otherwise, $T_a$ will be a zero vector.

\textbf{Order-2 INVT}: an order-2 INVT $T_{a_1a_2}$ has two indices which correspond to two sets of bases which carry the IRREPs $D_1(G)$ and $D_2(G)$. 
From the definition of INVT, one has $D_1(g)\, T \,D_2(g^{-1}) = T$ or equivalent $D_1(g)\, T = T \,D_2(g)$ for any $g\in G$. 
Then, by employing the famous Schur's lemma, one gets $T=0$ iff $D_1(G)$ and $D_2(G)$ are two nonequivalent IRREPs; 
for the case where $D_1(G)$ and $D_2(G)$ are equivalent, one gets $T$ must be the identity element of $G$.

\textbf{Order-3 INVT}: an order-3 INVT $T_{a_1a_2a_3}$ has three indices which correspond to three sets of baseas which carry the IRREPs $D_{a'_1}^{~~a_1}(G)$, $D_{a'_2}^{~~a_2}(G)$ and $D_{a'_3}^{~~a_3}(G)$, respectively. %
Here, an index $a_{23}$ can be defined as a pair combination of the indices $a_2$ and $a_3$ with the corresponding representation $D_{a'_{23}}^{~~a_{23}}(G) \equiv D_{a'_{2}}^{~~a_{2}}(G) \otimes D_{a'_{3}}^{~~a_{3}}(G)$. Then, $T_{a_1a_2a_3}$ can be expressed as an order-2 tensor $T_{a_1a_{23}}$.
In general, the representation $D_{a'_{23}}^{~~a_{23}}(G)$ is not an IRREP. Thus, it needs to be decomposed into a direct sum of some IRREPs, i.e., $D_{a'_{23}}^{~~a_{23}}(G) = D_{b'_{1}}^{~~b_{1}}(G) \oplus D_{b'_{2}}^{~~b_{2}}(G) \oplus \cdots$. With this decomposition, we can express the eq.~\eqref{eq:def_of_irten} of an order-3 INVT as a direct sum of eq.~\eqref{eq:def_of_irten} of many order-2 INVTs. 
Thus, we can apply Schur's lemma again: $T_{a_1a_2a_3}=0$ if any IRREP in the decomposition of $D_{a'_{23}}^{~~a_{23}}(G)$ is not equivalent to $D_{a'_1}^{~~a_1}(G)$; otherwise, $T_{a_1a_2a_3}$ must be uniquely determined by group transformations from eq.~\eqref{eq:def_of_irten}.

\textbf{Order-$n$ INVTs}: any order-$n$ ($n\geq4$) INVT can be expressed by contraction of order-3 INVTs.

Further, INVTs have a projective property that is independent of the bases and can be used to decompose COVTs. 
For example, consider an order-2 COVT $X_{ij}$ and direct product decomposition $[i]\otimes[j]=\left[k_1\right]\oplus\left[k_2\right]\oplus\cdots$, one has
\begin{equation}
	X_{ij}=C_{k_1}\, T^{k_1}_{ij} + C_{k_2}\, T^{k_2}_{ij} + \cdots,\quad \text{with}\quad C_{k_n} \xrightarrow{g\in G} \tilde{C}_{k_n} = D_{k_n}^{~~k_n'}(g)\,C_{k'_n},
	\label{eq:tensor_decompose}
\end{equation}
where $C_{k_n}~(n=1,2,\cdots)$ are a kind of sets of bases carrying IRREPs $\left[k_n\right]~(n=1,2,\cdots)$; $T^{k_n}_{ij}~(n=1,2,\cdots)$ are some order-3 INVTs.
These $C_{k_n}~(n=1,2,\cdots)$ can be obtained by the orthogonal relation of INVT as follows, 
\begin{align}
	T^{k_m}_{ij}~T_{k_n}^{ij} ~\propto~ \delta_{mn}~\delta_{k_n}^{~~k_m},
\end{align}
where the proportional sign represents the arbitrariness of normalization.
Eq.~\eqref{eq:tensor_decompose} clearly shows that a COVT can be decomposed into several parts by INVTs, while an INVT cannot be further decomposed. 
Thus, INVTs are also called IRTs.

From the above discussion, one can realize that CGCs are order-3 INVTs.
For example, it is easy to find that the CGCs of SU$(2$), $C^{sm}_{s_1m_1,s_2m_2}\equiv\left(C^{s}_{s_1s_2}\right)^{m}_{m_1m_2}$, are order-3 IRTs,
\begin{equation}
	\left(C^{s}_{s_1s_2}\right)^{m}_{m_1m_2}\xrightarrow{g\in\text{SU}(2)}D^{(s)m}_{m'}(g^{-1})~D^{(s_1)m_1'}_{m_1}(g)~D^{(s_2)m_2'}_{m_2}(g)~\left(C^{s}_{s_1s_2}\right)^{m'}_{m_1'm_2'}=\left(C^{s}_{s_1s_2}\right)^{m}_{m_1m_2},
	\label{eq:trans_nature_CGCs}
\end{equation}
where $D^{(s)m'}_{m}(g)$ is Wigner-$D$ matrix.
If one choose $s=1$, $s_1=\frac{1}{2}$ and $s_2=\frac{1}{2}$, the corresponding CGCs are $\left(C^{1}_{\frac{1}{2}\frac{1}{2}}\right)^{m}_{m_1m_2}$, where $m\in\{-1,0,1\}$ and $m_1,m_2\in\{-\frac{1}{2},\frac{1}{2}\}$. 
By using some similarity transformations to change the indices $(m,\,m_1,\,m_2)$ to $(i,\,a,\,b)$, the $\mathbf{2}\times\mathbf{2}$ matrices  $\left[\left(C^{1}_{\frac{1}{2}\frac{1}{2}}\right)^{m}\right]_{m_1m_2}$ will become Pauli-$\sigma$ matrices $(\sigma^i)_{a}^{~~b}~(a,b =1,2 ~\text{and}~ i=1,2,3)$.
Thus, Pauli-$\sigma$ matrices $(\sigma^i)_{a}^{~~b}$ form an order-3 IRT of SU$(2)$ with three indices $a,b$ and $i$ which carry three IRREPs $\left[\frac{1}{2}\right], \left[\frac{1}{2}\right]^*$ and $[1]$, respectively. Here, $[s]$ denotes the IRREP of SU(2) with dimension $2s+1$.

Actually, since the core concept of modern physics is symmetry, which is expressed as the principle of relativity --- the form of physical laws dose not depend on the selection of frame. 
Therefore, IRTs, a kind of invariant under group transformation, have become ideal mathematical objects to describe the laws of physics.
In particle physics, beside Pauli-$\sigma$ matrix, many common tensors are IRTs of specific group. 
For example, Minkowski metric $g_{\mu\nu}$, Levi-Civita tensor $\epsilon_{\mu\nu\rho\sigma}$ and Dirac-$\gamma$ matrix $\gamma_{\mu}$ are all IRTs of $L_p$.
Similarly, Dirac spinor wave functions $u^\sigma(\mathbf{p}), v^\sigma(\mathbf{p})$, and polarization vector $\epsilon_\mu^\sigma(\mathbf{p})$ are IRTs of the little group SO(3).
Furthermore, a particle's field operator is an IRT belonging to poinc$\acute{\text{a}}$re group --- a basic building block to construct poinc$\acute{\text{a}}$re invariant $S$-matrix.


\section{Partial wave components of some three-particle amplitudes}
\label{appd_LS_decomposition}

For a given Lorentz covariant three-particle amplitude, one can extract the partial wave amplitudes through IRTs and spin projection tensors as introduced in subsection~\ref{sec:irten_Lp}. Here, we provide a comprehensive list of all possible partial wave components for two spin-$\frac{1}{2}$ particles and for a spin-$\frac{1}{2}$ and a spin-$\frac{3}{2}$ particle using the aforementioned method. The results for two spin-$\frac{1}{2}$ particles can be found in table~\ref{tab:12_12_j_common_results}, while those for a spin-$\frac{1}{2}$ and a spin-$\frac{3}{2}$ particle can be found in table~\ref{tab:12_32_j_common_results}.

\begin{table}[htbp]
	\centering
	\begin{tabular}{c|c}
		\hline
		~~Lorentz structure~~& ~~ Partial wave components $\left({}^{2S+1}L_J\right)$ ~~ \\
		\hline
		\multirow{2}{*}{$\bar{\psi}_2\psi_1$} & ${}^3P_0$ \\
		& ${}^1S_0$ \\
		\hline
		\multirow{2}{*}{$\bar{\psi}_2\gamma_5\psi_1$}	& ${}^1S_0$ \\
		& ${}^3P_0$ \\
		\hline
		\multirow{2}{*}{$\bar{\psi}_2\gamma_\mu\psi_1$}	& $^3P_0~^3S_1~^3D_1$ \\
		& ${}^1S_0~{}^1P_1~{}^3P_1$ \\
		\hline
		\multirow{2}{*}{$\bar{\psi}_2\gamma_5\gamma_\mu\psi_1$}	&${}^3P_1~{}^1S_0~{}^1P_1$ \\
		& ${}^3P_0~{}^3S_1~{}^3D_1$ \\
		\hline
		\multirow{2}{*}{$\bar{\psi}_2\sigma_{\mu\nu}\psi_1$} & ${}^1P_1~{}^3S_1~{}^3P_1~{}^3D_1$ \\
		& ${}^1P_1~{}^3S_1~{}^3P_1~{}^3D_1$ \\
		\hline
		\multirow{2}{*}{$\bar{\psi}_2\stackrel{\leftrightarrow}{\partial}_{\mu}\psi_1$} & ${}^3S_1~{}^3P_0~{}^3D_1$ \\
		& ${}^1S_0$ \\
		\hline
		\multirow{2}{*}{$\partial_{\mu}\left(\bar{\psi}_2\psi_1\right)$} & ${}^3P_0$ \\
		& ${}^1S_0~{}^1P_1$ \\
		\hline
		\vdots & \vdots \\
		\hline
	\end{tabular}
\caption{Possible partial wave components contained in some common three-particle amplitudes including two fermions with spin-$\frac{1}{2}$.
	For each Lorentz structure, the upper row and lower row of the partial wave components represent the results of the decay mode $B\text{(boson)}\to F\text{(fermion)}\bar{F}\text{(antifermion)}$ and $F\to BF$, respectively.
}
	\label{tab:12_12_j_common_results}
\end{table}

\begin{table}[htbp]
	\centering
	\small
	\begin{tabular}{c|c}
		\hline
		~Lorentz structure~ & ~Partial wave components $\left({}^{2S+1}L_J\right)$~ \\
		\hline
		\multirow{2}{*}{$\bar{\psi}_2\psi_{1\mu}$} & $^5D_0~^3P_1~^5P_1~^5F_1$ \\
		& $^3P_0~^3S_1~^3D_1$ \\
		\hline
		\multirow{2}{*}{$\bar{\psi}_2\gamma_5\psi_{1\mu}$}	& $^3P_0~^3S_1~^3D_1$ \\
		& $^5D_0~^3P_1~^5P_1~^5F_1$ \\
		\hline
		\multirow{2}{*}{$\bar{\psi}_2\gamma_\nu\psi_{1\mu}$}	& $^5D_0~^3S_1~^3P_1~^5P_1~^3D_1~^5D_1~^5S_2~^3D_2~^5D_2$ \\
		& $^3P_0~^3S_1~^3P_1~^5P_1~^3D_1~^5D_1~^3P_2~^5P_2$ \\
		\hline
		\multirow{2}{*}{$\bar{\psi}_2\gamma_5\gamma_\nu\psi_{1\mu}$}	& $^3P_0~^3S_1~^3P_1~^5P_1~^3D_1~^5D_1~^3P_2~^5P_2$ \\
		& $^5D_0~^3S_1~^3P_1~^5P_1~^3D_1~^5D_1~^5S_2~^3D_2~^5D_2$ \\
		\hline
		\multirow{2}{*}{$\bar{\psi}_2\sigma_{\nu\rho}\psi_{1\mu}$} & ~$^3P_0~^5D_0~^3S_1~^3P_1~^5P_1~^3D_1~^5D_1~^5S_2~^3P_2~^5P_2~^3D_2~^5D_2$~\\
		& ~$^3P_0~^5D_0~^3S_1~^3P_1~^5P_1~^3D_1~^5D_1~^5S_2~^3P_2~^5P_2~^3D_2~^5D_2$~\\
		\hline
		\multirow{2}{*}{$\bar{\psi}_2\stackrel{\leftrightarrow}{\partial}_{\nu}\psi_{1\mu}$}	& $^5D_0~^3S_1~^3P_1~^5P_1~^3D_1~^5D_1~^5F_1~^5S_2~^3D_2~^5D_2~^5G_2$ \\
		& $^3P_0~^3S_1~^3D_1$ \\
		\hline
		\multirow{2}{*}{$\partial_{\nu}\left(\bar{\psi}_2\psi_{1\mu}\right)$} & $^5D_0~^3P_1~^5P_1~^5F_1$ \\
		& $^3P_0~^3S_1~^3P_1~^3D_1~^3P_2~^3F_2$ \\
		\hline
		\vdots & \vdots \\
		\hline
	\end{tabular}
	\caption{Notations are the same as in table~\ref{tab:12_12_j_common_results} except that a fermion is replaced by spin-$\frac{3}{2}$.
}
	\label{tab:12_32_j_common_results}
\end{table}


\section{Lorentz transformation in any irreducible representation of \boldmath{$L_p$}} 
\label{appd:Lp_transformation}

Since $L_p$ is a Lie group, a group element $g~(\in L_p)$ can be written as an exponential of its generators (table~\ref{tab:little_group}) as follows,
\begin{equation}
	g(\theta_i,\vartheta_i) = \exp{\left\{i\left( ~J_i~\theta_i +K_i~\vartheta_i~\right)\right\}},
\end{equation}
where $\theta_i$ and $\vartheta_i(i=1,2,3)$ are correspond to the group parameters of rotations and boosts, respectively. 
Equivalently, one can express $g$ by using $A_i$ and $B_i(i=1,2,3)$ (eq.~\eqref{eq:leftA_rightB}) as follows,
\begin{equation}
	g(z_i,z^*_i) = \exp{\left\{i\left( ~A_i~z^*_i +B_i~z_i~\right)\right\}},
	\label{eq:group_element_Lp_su2}
\end{equation}
where $z^*_i=\theta_i-i\vartheta_i(i=1,2,3)$ and $z_i=\theta_i+i\vartheta_i$ are corresponding to group parameters of the left and right-handed rotations, respectively. 
For an IRREP $(s_L,s_R)=(s_L,0)\otimes(0,s_R)$, the representation matrices of the generators will be $A^{(s_L,s_R)}_i = S_i^{[s_L]}\otimes\mathbbm{1}^{[s_R]}$ and $B^{(s_L,s_R)}_i = \mathbbm{1}^{[s_L]}\otimes S_i^{[s_R]}(i=1,2,3)$, where $S_i^{[s]}$ and $\mathbbm{1}^{[s]}$ are the SU(2) generators in IRREP $[s]$ and $(2s+1)$-dimensional identity matrix, respectively.
Thus, the representation of $L_p$ in arbitrary IRREP $[\alpha]=(s_L,s_R)$, denoted as $D_\alpha^{~~\alpha'}\left(\theta,\vartheta\right)$, is just a direct product of two Wigner-$D$ matrices with complex arguments as follows,
\begin{equation}
	D_\alpha^{~~\alpha'}\left(\theta_i,\vartheta_i\right) = U_\alpha^{lr}~U^{\alpha'}_{l'r'}~ D_{lr}^{~~l'r'}~\left(\theta_i,\vartheta_i\right) = U_\alpha^{lr}~U^{\alpha'}_{l'r'}~
	D_l^{(s_L)l'}(z^*_i)~D_r^{(s_R)r'}(z_i),
	\label{eq:appd_Lp_matrix_wigner_D}
\end{equation}
where $U_\alpha^{lr}$ and $U^\alpha_{lr}$ are defined in eq.~\eqref{eq:appd_index_map_alpha_to_lr}.
For a pure rotation, $z_i=z_i^*=\theta_i~(i=1,2,3)$ are reals, the corresponding transformation matrix is usual Wigner-$D$ matrix.
For a pure boost, $z_i=-z_i^*=i\vartheta_i~(i=1,2,3)$ are pure imaginary numbers.
Then, the corresponding transformation matrix can be obtained by performing the analytic continuation of usual Wigner-$D$ matrix.
According to eq.~\eqref{eq:convention_massive_h_p}, any pure boost transformation can be written as follows,
\begin{equation}
	D_\alpha^{~~\alpha'}\left(h_{\mathbf{p}}\right) = D_\alpha^{~~\alpha''}\left(R_{\hat{\mathbf{p}}}\right)~D_{\alpha''}^{~~\alpha'''}\left(B_{|\mathbf{p}|}\right)~D_{\alpha'''}^{~~\alpha'}\left(R_{\hat{\mathbf{p}}}^{-1}\right),
	\label{eq:appd_pure_boost_any_irrep}
\end{equation}
where
\begin{align}
	D_{\alpha}^{~~\alpha'}\left(\,R_{\hat{\mathbf{p}}}\,\right) &= U_\alpha^{lr}~U^{\alpha'}_{l'r'}~D_l^{(s_L)l''}(\theta_3)~D_{l''}^{(s_L)l'}(\theta_2)~D_r^{(s_R)r''}(\theta_3)~D_{r''}^{(s_R)r'}(\theta_2),\notag\\
	D_{\alpha}^{~~\alpha'}\left(B_{|\mathbf{p}|}\right) &= U_\alpha^{lr}~U^{\alpha'}_{l'r'}~D_l^{(s_L)l'}(-i\vartheta_3)~D_r^{(s_R)r'}(i\vartheta_3),\notag
\end{align}
with $D_\sigma^{(s)\sigma'}(\pm i \vartheta_3) = \delta_\sigma^{~~\sigma'} e^{\mp \sigma \vartheta_3}$.

Finally, one can  obtain a Lorentz transformation in arbitrary representation from eq.~\eqref{eq:appd_pure_boost_any_irrep}.
For example, considering the fundamental representation $[\alpha]=(\frac{1}{2},\frac{1}{2})$, one obtains
\begin{align}
	D_\mu^{~~\mu'}\left(h_\mathbf{p}\right) =&~ U_\mu^{~~\alpha}~D_\alpha^{~~\alpha'}\left(h_{\mathbf{p}}\right)~\left(U^{-1}\right)^{~~\mu'}_{\alpha'}
	\notag\\
	=&
	\left(\begin{array}{cc}
		\operatorname{cosh}[\vartheta_3] & \operatorname{cos}[\theta_3] \operatorname{sin}[\theta_2] \operatorname{sinh}[\vartheta_3] \\
		~~\operatorname{cos}[\theta_3] \operatorname{sin}[\theta_2] \operatorname{sinh}[\vartheta_3]~~ & \operatorname{cos}[\theta_3]^{2}\left(\operatorname{cos}[\theta_2]^{2}+\operatorname{cosh}[\vartheta_3] \operatorname{sin}[\theta_2]^{2}\right)+\operatorname{sin}[\theta_3]^{2} \\
		\operatorname{sin}[\theta_2] \operatorname{sin}[\theta_3] \operatorname{sinh}[\vartheta_3] & \operatorname{sin}[\theta_2]^{2} \operatorname{sin}[2 \theta_3] \operatorname{sinh}\left[\frac{\vartheta_3}{2}\right]^{2} \\
		\operatorname{cos}[\theta_2] \operatorname{sinh}[\vartheta_3] & \operatorname{sin}[2\theta_2] \operatorname{cos}[\theta_3]\operatorname{sinh}\left[\frac{\vartheta_3}{2}\right]^{2}
	\end{array}\right.\notag\\
	&\hspace{-1cm}\left.\begin{array}{cc}
		\operatorname{sin}[\theta_2] \operatorname{sin}[\theta_3] \operatorname{sinh}[\vartheta_3] & \operatorname{cos}[\theta_2] \operatorname{sinh}[\vartheta_3] \\
		\operatorname{sin}[\theta_2]^{2} \operatorname{sin}[2 \theta_3] \operatorname{sinh}\left[\frac{\vartheta_3}{2}\right]^{2} & \operatorname{sin}[2\theta_2] \operatorname{cos}[\theta_3]\operatorname{sinh}\left[\frac{\vartheta_3}{2}\right]^{2} \\
		\operatorname{cos}[\theta_3]^{2}+\left(\operatorname{cos}[\theta_2]^{2}+\operatorname{cosh}[\vartheta_3] \operatorname{sin}[\theta_2]^{2}\right) \operatorname{sin}[\theta_3]^{2} & \operatorname{sin}[2\theta_2] \operatorname{sin}[\theta_3]\operatorname{sinh}\left[\frac{\vartheta_3}{2}\right]^{2} \\
		\operatorname{sin}[2\theta_2] \operatorname{sin}[\theta_3]\operatorname{sinh}\left[\frac{\vartheta_3}{2}\right]^{2} & ~\operatorname{cos}[\theta_2]^{2} \operatorname{cosh}[\vartheta_3]+\operatorname{sin}[\theta_2]^{2}~
	\end{array}\right)_\mu^{~~\mu'},
	\label{eq:appd_Lorentz_trans_explicit}
\end{align}
where $U_\mu^{~~\alpha}$ and $\left(U^{-1}\right)^{~~\mu}_{\alpha}$ are similarity transformations, defined as follows,
\begin{equation}
	U_\mu^{~~\alpha} ~=~ \left(\begin{array}{cccc}
		0 & \frac{1}{\sqrt{2}} & -\frac{1}{\sqrt{2}} & 0 \\
		\frac{1}{\sqrt{2}} & 0 & 0 & -\frac{1}{\sqrt{2}} \\
		\frac{i}{\sqrt{2}} & 0 & 0 & \frac{i}{\sqrt{2}} \\
		0 & -\frac{1}{\sqrt{2}} & -\frac{1}{\sqrt{2}} & 0
	\end{array}\right)_\mu^{~~\alpha},\qquad
	\left(U^{-1}\right)_{\alpha}^{~~\mu} ~=~ \left(\begin{array}{cccc}
		0 & \frac{1}{\sqrt{2}} & -\frac{i}{\sqrt{2}} & 0 \\
		\frac{1}{\sqrt{2}} & 0 & 0 & -\frac{1}{\sqrt{2}} \\
		-\frac{1}{\sqrt{2}} & 0 & 0 & -\frac{1}{\sqrt{2}} \\
		0 & -\frac{1}{\sqrt{2}} & -\frac{i}{\sqrt{2}} & 0
	\end{array}\right)_{\alpha}^{~~\mu}.
	\label{eq:appd_spacetime_chiral}
\end{equation}


\section{Some examples of deriving irreducible tensors of \boldmath{$L_p$}}
\label{appd:ex_irten_Lp}

In this appendix, we present examples on how to derive IRTs of $L_p$. The fundamental representation $\left(\frac{1}{2},\frac{1}{2}\right)$ can be expressed as $\left(\frac{1}{2},0\right)\otimes\left(0,\frac{1}{2}\right)$, which implies the existence of an order-3 IRT, as shown below,
\begin{equation}
	\left(\frac{1}{2},\frac{1}{2}\right) ~\mapsto~ \left(\frac{1}{2},0\right)\otimes\left(0,\frac{1}{2}\right) ~:~~ T_{lr}^{\mu},
\end{equation}
where we adopt the conventions presented in table~\ref{tab:symbol_conventions} for simplicity (these conventions will be used throughout the following without declaration).
According to eq.~\eqref{eq:irten_order_3_CGCs}, one has 
\begin{equation}
	T_{lr}^{\mu} ~=~ \sum_{s} u^{\sigma}_{lr}\left(\mathbf{k};\left[\left(\frac{1}{2},0\right)\otimes\left(0,\frac{1}{2}\right)\right],s\right) \bar{u}^{\mu}_{\sigma}\left(\mathbf{k};\left(\frac{1}{2},\frac{1}{2}\right),s\right),
	\label{eq:irten_1212_12x12}
\end{equation}
and by using eqs.~\eqref{eq:spin_wave_func_CGCs} and \eqref{eq:spin_wave_func_direct_product}, one can obtain 
\begin{equation}
	T_{lr}^{\mu} ~=~\sum_{s} \left(C^s_{\frac{1}{2}\frac{1}{2}}\right)^\sigma_{lr}~\left(C_s^{\frac{1}{2}\frac{1}{2}}\right)_\sigma^{l'r'}~U^\mu_{~~\alpha}~U^\alpha_{l'r'}~=\left(U^{-1}\right)^{~~\mu}_{\alpha}~U^\alpha_{lr}.
	\label{eq:irten_1212_12x12_similarity_trans}
\end{equation}
where $U^\alpha_{lr}$ and $\left(U^{-1}\right)^{~~\mu}_{\alpha}$ are specified in eqs.~\eqref{eq:appd_index_map_alpha_to_lr} and \eqref{eq:appd_spacetime_chiral}, respectively.
Therefore, the order-3 IRT $T_{lr}^{\mu}$ is just a similarity transformation, which is known as four-dimensional Pauli matrix  $\left(\sigma^\mu\right)_l^{~~r}$.

Next, we discuss the direct product decomposition of two $(\frac{1}{2},\frac{1}{2})$ IRREPs, which can be written as follows,
\begin{equation}
	\left(\frac{1}{2},\frac{1}{2}\right)\otimes\left(\frac{1}{2},\frac{1}{2}\right)=\left(0,0\right)\oplus\left(1,0\right)\oplus\left(0,1\right)\oplus\left(1,1\right).
	\label{eq:direct_prod_decomposition_1212x1212}
\end{equation}
This decomposition gives rise to four IRREPs and their corresponding IRTs can be expressed as
\begin{align}
	\left(\frac{1}{2},\frac{1}{2}\right)\otimes\left(\frac{1}{2},\frac{1}{2}\right)~&\mapsto~(0,0) ~:~~T^{\mu\nu}_{\mu^0},\qquad
	\left(\frac{1}{2},\frac{1}{2}\right)\otimes\left(\frac{1}{2},\frac{1}{2}\right)~\mapsto~(1,0) ~:~~ T^{\mu\nu}_{l},\notag\\
	\left(\frac{1}{2},\frac{1}{2}\right)\otimes\left(\frac{1}{2},\frac{1}{2}\right)~&\mapsto~(0,1) ~:~~ T^{\mu\nu}_{r},\qquad
	\left(\frac{1}{2},\frac{1}{2}\right)\otimes\left(\frac{1}{2},\frac{1}{2}\right)~\mapsto~(1,1) ~:~~ T^{\mu\nu}_{\mu^2}.
\end{align}
Using eq.~\eqref{eq:irten_order_3_CGCs}, we obtain
\begin{align}
	T^{\mu\nu}_{\mu^0} ~&=~U^{00}_{\mu^0} \left(C_0^{\frac{1}{2}\frac{1}{2}}\right)_0^{l_1l_2}\left(C_0^{\frac{1}{2}\frac{1}{2}}\right)_0^{r_1r_2} T^\mu_{l_1r_1}T^\nu_{l_2r_2},\quad
	T^{\mu\nu}_l ~=~ \left(C_1^{\frac{1}{2}\frac{1}{2}}\right)_{l}^{l_1l_2}\left(C_0^{\frac{1}{2}\frac{1}{2}}\right)_{0}^{r_1r_2} T^\mu_{l_1r_1}T^\nu_{l_2r_2},\notag\\
	T^{\mu\nu}_{\mu^2} ~&=~ U^{lr}_{\mu^2}\left(C_1^{\frac{1}{2}\frac{1}{2}}\right)_{l}^{l_1l_2}\left(C_1^{\frac{1}{2}\frac{1}{2}}\right)_{r}^{r_1r_2}T^\mu_{l_1r_1}T^\nu_{l_2r_2},\quad
	T^{\mu\nu}_r ~=~ \left(C_0^{\frac{1}{2}\frac{1}{2}}\right)_{0}^{l_1l_2}\left(C_1^{\frac{1}{2}\frac{1}{2}}\right)_{r}^{r_1r_2} T^\mu_{l_1r_1}T^\nu_{l_2r_2},
	\label{eq:irten_1212x1212}
\end{align}
where $T^\mu_{lr}$ is given by eq.~\eqref{eq:irten_1212_12x12_similarity_trans}; 
$U^{00}_{\mu^0}$ and $U^{lr}_{\mu^2}$ can be obtained from eq.~\eqref{eq:appd_index_map_alpha_to_lr}.
The IRTs in eq.~\eqref{eq:irten_1212x1212} can be expressed in a familiar way, containing only Lorentz four-vector indices, as follows,
\begin{align}
	\left(T_{(0,0)}\right)^{\mu\nu,\mu'\nu'} ~&\equiv~ T^{\mu\nu}_{\mu^0}  T^{\mu^0\mu'\nu'} ~=~ \frac{1}{4}~g^{\mu\nu}g^{\mu'\nu'},\notag\\
	\left(T_{\left[(1,0)\oplus(0,1)\right]^+}\right)^{\mu\nu,\mu'\nu'} ~&\equiv~  T^{\mu\nu}_l T^{l\mu'\nu'} + T^{\mu\nu}_r T^{r\mu'\nu'} ~=~ \frac{1}{2}~\left(g^{\mu\mu'}g^{\nu\nu'}-g^{\mu\nu'}g^{\nu\mu'}\right),\notag\\
	\left(T_{\left[(1,0)\oplus(0,1)\right]^-}\right)^{\mu\nu,\mu'\nu'} ~&\equiv~  T^{\mu\nu}_l T^{l\mu'\nu'} - T^{\mu\nu}_r T^{r\mu'\nu'} ~=~ \frac{i}{2}~ \epsilon^{\mu\nu\mu'\nu'},\notag\\
	\left(T_{(1,1)}\right)^{\mu\nu,\mu'\nu'} ~&\equiv~ T^{\mu\nu}_{\mu^2}  T^{\mu^2\mu'\nu'} ~=~ \frac{1}{2}~\left(g^{\mu\mu'}g^{\nu\nu'}+g^{\mu\nu'}g^{\nu\mu'}\right)-\frac{1}{4}~g^{\mu\nu}g^{\mu'\nu'}.
	\label{eq:irten_1212x1212_Lorentz}
\end{align}

Then, we discuss two examples with Dirac spinor representation $\left[\frac{1}{2},0\right]$. 
The direct product decomposition of two $\left[\frac{1}{2},0\right]$ representations is as follows,
\begin{align}
	\left[\left(\frac{1}{2},0\right)\oplus\left(0,\frac{1}{2}\right)\right]\otimes&\left[\left(\frac{1}{2},0\right)\oplus\left(0,\frac{1}{2}\right)\right] ~=~\notag\\
	&\left(0,0\right)_L\oplus\left(0,0\right)_R\oplus\left(1,0\right)\oplus\left(0,1\right)\oplus\left(\frac{1}{2},\frac{1}{2}\right)_L\oplus\left(\frac{1}{2},\frac{1}{2}\right)_R.
	\label{eq:12_12_decompose}
\end{align}
This yields six corresponding IRTs as follows,
\begin{align}
	\left[\left(\frac{1}{2},0\right)\oplus\left(0,\frac{1}{2}\right)\right]\otimes\left[\left(\frac{1}{2},0\right)\oplus\left(0,\frac{1}{2}\right)\right]~&\mapsto~(0,0)_L ~:~~\left(T_L\right)^{ab}_{\mu^0},\notag\\
	\left[\left(\frac{1}{2},0\right)\oplus\left(0,\frac{1}{2}\right)\right]\otimes\left[\left(\frac{1}{2},0\right)\oplus\left(0,\frac{1}{2}\right)\right]~&\mapsto~(0,0)_R ~:~~\left(T_R\right)^{ab}_{\mu^0},\notag\\
	\left[\left(\frac{1}{2},0\right)\oplus\left(0,\frac{1}{2}\right)\right]\otimes\left[\left(\frac{1}{2},0\right)\oplus\left(0,\frac{1}{2}\right)\right]~&\mapsto~(1,0) ~:~~ T^{ab}_{l},\notag\\
	\left[\left(\frac{1}{2},0\right)\oplus\left(0,\frac{1}{2}\right)\right]\otimes\left[\left(\frac{1}{2},0\right)\oplus\left(0,\frac{1}{2}\right)\right]~&\mapsto~(0,1) ~:~~ T^{ab}_{r},\notag\\
	\left[\left(\frac{1}{2},0\right)\oplus\left(0,\frac{1}{2}\right)\right]\otimes\left[\left(\frac{1}{2},0\right)\oplus\left(0,\frac{1}{2}\right)\right]~&\mapsto~\left(\frac{1}{2},\frac{1}{2}\right)_L ~:~~ \left(T_L\right)^{ab}_{\mu},\notag\\
	\left[\left(\frac{1}{2},0\right)\oplus\left(0,\frac{1}{2}\right)\right]\otimes\left[\left(\frac{1}{2},0\right)\oplus\left(0,\frac{1}{2}\right)\right]~&\mapsto~\left(\frac{1}{2},\frac{1}{2}\right)_R ~:~~ \left(T_R\right)^{ab}_{\mu}.
\end{align}
According to eq.~\eqref{eq:irten_order_3_CGCs}, we obtain
\begin{align}
	\left(T_L\right)^{ab}_{\mu^0} ~&=~   \left(C^{\frac{1}{2}\frac{1}{2}}_0\right)^{l_1l_2}_{0}\left(C^{00}_0\right)^{r_1r_2}_{0} \left(U_L\right)^a_{l_1r_1}\left(U_L\right)^b_{l_2r_2}
	~\equiv~ \left(T_L\right)^{ab},\notag\\
	\left(T_R\right)^{ab}_{\mu^0} ~&=~ \left(C^{00}_0\right)^{l_1l_2}_{0}\left(C^{\frac{1}{2}\frac{1}{2}}_0\right)^{r_1r_2}_{0}\left(U_R\right)^a_{l_1r_1}\left(U_R\right)^b_{l_2r_2}
	~\equiv~ \left(T_R\right)^{ab},\notag\\
	T^{ab}_l 
	~&=~\left(C^{\frac{1}{2}\frac{1}{2}}_1\right)^{l_1l_2}_l\left(C^{00}_0\right)^{r_1r_2}_0 \left(U_L\right)^a_{l_1r_1}\left(U_L\right)^b_{l_2r_2},\notag\\
	T^{ab}_r 
	~&=~\left(C^{00}_0\right)^{l_1l_2}_0\left(C^{\frac{1}{2}\frac{1}{2}}_1\right)^{r_1r_2}_r \left(U_R\right)^a_{l_1r_1}\left(U_R\right)^b_{l_2r_2},\notag\\
	\left(T_L\right)^{ab}_{\mu} 
	~&=~ \left(C^{\frac{1}{2}0}_{\frac{1}{2}}\right)^{l_1l_2}_{l}\left(C^{0\frac{1}{2}}_{\frac{1}{2}}\right)^{r_1r_2}_{r} T^{lr}_\mu \left(U_L\right)^a_{l_1r_1}\left(U_R\right)^b_{l_2r_2},\notag\\
	\left(T_R\right)^{ab}_{\mu} 
	~&=~ \left(C^{0\frac{1}{2}}_{\frac{1}{2}}\right)^{l_1l_2}_{l}\left(C^{\frac{1}{2}0}_{\frac{1}{2}}\right)^{r_1r_2}_{r} T^{lr}_\mu \left(U_R\right)^a_{l_1r_1}\left(U_L\right)^b_{l_2r_2},
	\label{eq:irten_12+12x12+12}
\end{align}
where $T^{lr}_\mu$ is the same as that in eq.~\eqref{eq:irten_1212_12x12_similarity_trans}; 
$\left(U_{L/R}\right)^a_{lr}$ can be obtained from eq.~\eqref{eq:appd_index_map_alpha_to_lr_rerep}.
The IRTs in eq.~\eqref{eq:irten_12+12x12+12} can be expressed in a familiar way (only contains Lorentz four-vector indices and Dirac spinor indices) as follows,
\begin{align}
	\left(T_{(0,0)^+}\right)^{ab} ~&\equiv~ \left(T_L\right)^{ab}+\left(T_R\right)^{ab} ~\equiv~ g^{ab},\notag\\
	\left(T_{(0,0)^-}\right)^{ab} ~&\equiv~ \left(T_L\right)^{ab}-\left(T_R\right)^{ab} ~=~ g^{ac} \left(\gamma_5\right)_c^{~~b},\notag\\
	\left(T_{[1,0]^+}\right)^{ab,\mu\nu} ~&\equiv~ T^{ab}_l\, T^{l\mu\nu} + T^{ab}_r\, T^{r\mu\nu} ~=~ \frac{-i}{\sqrt{2}} g^{ac}\left(\sigma^{\mu\nu}\right)_c^{~~b},\notag\\
	\left(T_{[1,0]^-}\right)^{ab,\mu\nu} ~&\equiv~ T^{ab}_l\, T^{l\mu\nu} - T^{ab}_r\, T^{r\mu\nu} ~=~ \frac{-i}{\sqrt{2}} g^{ac}\left(\gamma_5 \sigma^{\mu\nu}\right)_c^{~~b},\notag\\
	\left(T_{\left(\frac{1}{2},\frac{1}{2}\right)^+}\right)^{ab}_\mu ~&\equiv~ \left(T_L\right)^{ab}_{\mu}+\left(T_R\right)^{ab}_{\mu} ~=~ \frac{1}{\sqrt{2}} g^{ac} \left(\gamma_5 \gamma_\mu\right)_c^{~~b},\notag\\
	\left(T_{\left(\frac{1}{2},\frac{1}{2}\right)^-}\right)^{ab}_\mu ~&\equiv~ \left(T_L\right)^{ab}_{\mu}-\left(T_R\right)^{ab}_{\mu} ~=~ \frac{1}{\sqrt{2}} g^{ac} \left(\gamma_\mu\right)_c^{~~b},
	\label{eq:irten_12+12x12+12_Lorentz_Dirac}
\end{align}
where $\gamma^\mu$ and $\gamma_5$ are Dirac-$\gamma$ matrices; 
$\sigma^{\mu\nu}=\frac{i}{2}\left[\gamma^\mu,\gamma^\nu\right]$; 
$g^{ab}$ is the metric of Dirac spinor space, with the following explicit form,
\begin{equation}
	g^{ab}=\begin{pmatrix}
		0 & ~1~ & 0 & 0 \\
		-1~ & 0 & 0 & 0 \\
		0 & 0 & 0 & ~1~ \\
		0 & 0 & -1~ & 0 
	\end{pmatrix}^{ab}, \qquad g_{ab}=-g^{ab}.
\end{equation}

We then consider the direct product decomposition of Dirac spinor representation $\left[\frac{1}{2},0\right]$ and fundamental representation $\left(\frac{1}{2},\frac{1}{2}\right)$ as follows,
\begin{equation}
	\left[\left(\frac{1}{2},0\right)\oplus\left(0,\frac{1}{2}\right)\right]\otimes\left(\frac{1}{2},\frac{1}{2}\right) = \left(\frac{1}{2},0\right)\oplus\left(0,\frac{1}{2}\right)\oplus\left(1,\frac{1}{2}\right)\oplus\left(\frac{1}{2},1\right).
\end{equation}
This yields four corresponding IRTs as follows,
\begin{align}
	\left[\left(\frac{1}{2},0\right)\oplus\left(0,\frac{1}{2}\right)\right]\otimes\left(\frac{1}{2},\frac{1}{2}\right)~&\mapsto~\left(\frac{1}{2},0\right) ~:~~T^{a\mu}_{l},\notag\\
	\left[\left(\frac{1}{2},0\right)\oplus\left(0,\frac{1}{2}\right)\right]\otimes\left(\frac{1}{2},\frac{1}{2}\right)~&\mapsto~\left(0,\frac{1}{2}\right) ~:~~T^{a\mu}_{r},\notag\\
	\left[\left(\frac{1}{2},0\right)\oplus\left(0,\frac{1}{2}\right)\right]\otimes\left(\frac{1}{2},\frac{1}{2}\right)~&\mapsto~\left(1,\frac{1}{2}\right) ~:~~ \left(T_L\right)^{a\mu}_{a^3},\notag\\
	\left[\left(\frac{1}{2},0\right)\oplus\left(0,\frac{1}{2}\right)\right]\otimes\left(\frac{1}{2},\frac{1}{2}\right)~&\mapsto~\left(\frac{1}{2},1\right) ~:~~ \left(T_R\right)^{a\mu}_{a^3}.
\end{align}
Using eq.~\eqref{eq:irten_order_3_CGCs}, we have
\begin{align}
	T^{a\mu}_{l} 
	~&=~\left(C^{0\frac{1}{2}}_\frac{1}{2}\right)^{l_1l_2}_{l}\left(C^{\frac{1}{2}\frac{1}{2}}_0\right)^{r_1r_2}_{0}\left(U_R\right)^a_{l_1r_1}T^\mu_{l_2r_2},\notag\\
	T^{a\mu}_{r} 
	~&=~\left(C^{\frac{1}{2}\frac{1}{2}}_0\right)^{l_1l_2}_{0}\left(C^{0\frac{1}{2}}_\frac{1}{2}\right)^{r_1r_2}_{r}\left(U_L\right)^a_{l_1r_1}T^\mu_{l_2r_2},\notag\\
	\left(T_L\right)^{a\mu}_{a^3} 
	~&=~\left(C^{\frac{1}{2}\frac{1}{2}}_1\right)^{l_1l_2}_{l}\left(C^{\frac{1}{2}0}_\frac{1}{2}\right)^{r_1r_2}_{r}\left(U_L\right)^{lr}_{a^3}\left(U_L\right)^a_{l_1r_1}T^\mu_{l_2r_2},\notag\\
	\left(T_R\right)^{a\mu}_{a^3}
	~&=~\left(C^{0\frac{1}{2}}_\frac{1}{2}\right)^{l_1l_2}_{l}\left(C^{\frac{1}{2}\frac{1}{2}}_1\right)^{r_1r_2}_{r}\left(U_R\right)^{lr}_{a^3}\left(U_R\right)^a_{l_1r_1}T^\mu_{l_2r_2},
	\label{eq:irten_12+12x1212}
\end{align}
where $T^\mu_{lr}$ is the same as that in eq.~\eqref{eq:irten_1212_12x12_similarity_trans}; 
$\left(U_{L/R}\right)^a_{lr}$ and $\left(U_{L/R}\right)^{lr}_{a^3}$ can be obtained from eq.~\eqref{eq:appd_index_map_alpha_to_lr_rerep}.
Similarly, the IRTs in eq.~\eqref{eq:irten_12+12x1212} can be expressed in a familiar way (only contains Lorentz four-vector indices and Dirac spinor indices) as follows,
\begin{align}
	\left(T_{\left[\frac{1}{2},0\right]^+}\right)^{a\mu}_{b} ~&=~ \left(U_L\right)^{l0}_b\, T^{a\mu}_l + \left(U_R\right)^{0r}_b\, T^{a\mu}_r ~=~ \left(\gamma_5\gamma^\mu\right)_b^{~~a},\notag\\
	\left(T_{\left[\frac{1}{2},0\right]^-}\right)^{a\mu}_{b} ~&=~ \left(U_L\right)^{l0}_b\, T^{a\mu}_l - \left(U_R\right)^{0r}_b\, T^{a\mu}_r ~=~ \left(\gamma^\mu\right)_b^{~~a},\notag\\
	\left(T_{\left[1,\frac{1}{2}\right]^+}\right)^{a\mu}_{b\nu} ~&=~ \left(T_L\right)_{b\nu}^{a^3}\, \left(T_L\right)^{a\mu}_{a^3} + \left(T_R\right)_{b\nu}^{a^3}\, \left(T_R\right)^{a\mu}_{a^3}
	~=~ \frac{3}{4}\, g^{~~\mu}_\nu\, \delta_b^{~~a} - \frac{i}{4}\, g^{\mu\rho}\, \left(\sigma_{\rho\nu}\right)_b^{~~a},\notag\\
	\left(T_{\left[1,\frac{1}{2}\right]^-}\right)^{a\mu}_{b\nu} ~&=~ \left(T_L\right)_{b\nu}^{a^3}\, \left(T_L\right)^{a\mu}_{a^3} - \left(T_R\right)_{b\nu}^{a^3}\, \left(T_R\right)^{a\mu}_{a^3}
	~=~ \frac{3}{4}\, g^{~~\mu}_\nu\, \left(\gamma_5\right)_b^{~~a} - \frac{i}{4}\, g^{\mu\rho}\, \left(\gamma_5\sigma_{\rho\nu}\right)_b^{~~a},
\end{align}
where $\left(U_{L/R}\right)^{lr}_b$ is the same as that in eq.~\eqref{eq:irten_12+12x12+12} with $r=0$ or $l=0$.

The derivation of any other IRT of $L_p$ is similar.
These IRTs serve as the building blocks for constructing Lorentz covariant and invariant structures.

\section{Some examples of deriving irreducible tensors of little group SO(3)}
\label{appd:ex_irten_so3}

In this appendix, we present examples of IRTs of the little group SO(3).
Firstly, for a specific class of the spin projection tensors with $[\beta]=[\alpha_1]\equiv[\alpha]$ and $[\alpha_2]=(0,0)$ in eq.~\eqref{eq:irten_little_group_so3}, one has the following spin projection tensor,
\begin{equation}
	P_\alpha^{~~\beta}(\mathbf{p};\chi_{_1},\chi_{_2},s) ~=~  u_\alpha^{\sigma}(\mathbf{p};\chi_{_1},s)\,\bar{u}_{\sigma}^{\beta}(\mathbf{p};\chi_{_2}^*,s).
	\label{eq:irten_so3_proj_tensor}
\end{equation}
Using the orthonormal relation in eq.~\eqref{eq:self_conj_rep_oorthogonal_norm}, we obtain
\begin{equation}
	P_\alpha^{~~\beta}(\mathbf{p};\chi_{_1},\chi_{_2},s_1)\, u_{\beta}^{\sigma}(\mathbf{p};\chi_{_1},s_2) ~=~ \delta_{\chi_{_1}\chi_{_2}}\, \delta_{s_1s_2}\, u_{\alpha}^{\sigma}(\mathbf{p};\chi_{_1},s_1),
\end{equation}
which is the relativistic equation of spin wave functions in momentum space.
For instance, considering the case of $[\alpha]=\left(\frac{1}{2},\frac{1}{2}\right)$, one has the following spin projection tensors,
\begin{align}
	P_{\mu}^{~~\nu}\left(\mathbf{p};\left(\frac{1}{2},\frac{1}{2}\right)\right.,\left.\left(\frac{1}{2},\frac{1}{2}\right),0\right)
	=~& \frac{p_{\mu}p^{\nu}}{p^2},\notag\\
	P_{\mu}^{~~\nu}\left(\mathbf{p};\left(\frac{1}{2},\frac{1}{2}\right)\right.,\left.\left(\frac{1}{2},\frac{1}{2}\right),1\right)
	=~& g^{~~\nu}_{\mu}-\frac{p_{\mu}p^{\nu}}{p^2}.
	\label{eq:proj_spin_tensor_1212_01}
\end{align}
The two spin projection tensors in eq.~\eqref{eq:proj_spin_tensor_1212_01} implies Proca equation for a massive vector particle, i.e., $\partial_\mu F^{\mu\nu}+m^2 A^\nu=0$ with $F^{\mu\nu}=\partial^\mu A^\nu-\partial^\nu A^\mu$. This equation reduces to the Maxwell equation when $m=0$.

For a self-conjugate REREP $[s_L,s_R]~(s_L\neq s_R)$, one can transform the left and right-handed states $u_{\alpha}^{\sigma}(\mathbf{p};(s_L,s_R),s)$ and $u_{\alpha}^{\sigma}(\mathbf{p};(s_R,s_L),s)$ as shown in eq.~\eqref{eq:self_conj_rep_spin_wave_func} into parity eigenstates as follows,
\begin{align}
	\text{Positive parity~:}\quad	u_{\alpha}^{\sigma}(\mathbf{p};[s_L,s_R],s^+) ~&=~ \frac{1}{\sqrt{2}}\left[u_{\alpha}^{\sigma}(\mathbf{p};(s_L,s_R),s) +  u_{\alpha}^{\sigma}(\mathbf{p};(s_R,s_L),s)\right],\notag\\
	\bar{u}^{\alpha}_{\sigma}(\mathbf{p};[s_L,s_R],s^+) ~&=~ \frac{1}{\sqrt{2}}\left[\bar{u}^{\alpha}_{\sigma}(\mathbf{p};(s_L,s_R),s) +  \bar{u}^{\alpha}_{\sigma}(\mathbf{p};(s_R,s_L),s)\right],\notag\\
	\text{Negative parity~:}\quad
	u_{\alpha}^{\sigma}(\mathbf{p};[s_L,s_R],s^-) ~&=~ \frac{1}{\sqrt{2}}\left[u_{\alpha}^{\sigma}(\mathbf{p};(s_L,s_R),s) -  u_{\alpha}^{\sigma}(\mathbf{p};(s_R,s_L),s)\right],\notag\\
	\bar{u}^{\alpha}_{\sigma}(\mathbf{p};[s_L,s_R],s^-) ~&=~ \frac{1}{\sqrt{2}}\left[\bar{u}^{\alpha}_{\sigma}(\mathbf{p};(s_R,s_L),s) -  \bar{u}^{\alpha}_{\sigma}(\mathbf{p};(s_L,s_R),s)\right].
	\label{eq:spin_wave_func_parity_eigenstate}
\end{align}
Using this transformation, one can obtain the relativistic equations for particle with any half-integer spin and definite parity as follows,
\begin{align}
	P_\alpha^{~~\beta}(\mathbf{p};[s_L,s_R],s^+)\, u_{\beta}^{\sigma}(\mathbf{p};[s_L,s_R],s^+) ~=&~ u_{\alpha}^{\sigma}(\mathbf{p};[s_L,s_R],s^+),\notag\\
	P_\alpha^{~~\beta}(\mathbf{p};[s_L,s_R],s^+)\, u_{\beta}^{\sigma}(\mathbf{p};[s_L,s_R],s^-) ~=&~ 0,\notag\\
	P_\alpha^{~~\beta}(\mathbf{p};[s_L,s_R],s^-)\, u_{\beta}^{\sigma}(\mathbf{p};[s_L,s_R],s^-) ~=&~ u_{\alpha}^{\sigma}(\mathbf{p};[s_L,s_R],s^-),\notag\\
	P_\alpha^{~~\beta}(\mathbf{p};[s_L,s_R],s^-)\, u_{\beta}^{\sigma}(\mathbf{p};[s_L,s_R],s^+) ~=&~ 0,
\end{align}
where
\begin{align}
	P_\alpha^{~~\beta}(\mathbf{p};[s_L,s_R],s^+) ~=&~  u_{\alpha}^{\sigma}(\mathbf{p};[s_L,s_R],s^+) \bar{u}^{\beta}_{\sigma}(\mathbf{p};[s_L,s_R],s^+),\notag\\
	P_\alpha^{~~\beta}(\mathbf{p};[s_L,s_R],s^-) ~=&~-  u_{\alpha}^{\sigma}(\mathbf{p};[s_L,s_R],s^-) \bar{u}^{\beta}_{\sigma}(\mathbf{p};[s_L,s_R],s^-).
\end{align}
For instance, considering the case of $[s_L,s_R]=\left[\frac{1}{2},0\right]$, one will get
\begin{align}
	P_a^{~~b}\left(\mathbf{p};\left[\frac{1}{2},0\right],\frac{1}{2}^+\right) ~&=~  u_{a}^{\sigma}\left(\mathbf{p};\left[\frac{1}{2},0\right],\frac{1}{2}^+\right) \bar{u}^{b}_{\sigma}\left(\mathbf{p};\left[\frac{1}{2},0\right],\frac{1}{2}^+\right),\notag\\
	P_a^{~~b}\left(\mathbf{p};\left[\frac{1}{2},0\right],\frac{1}{2}^-\right) ~&=~ -\, u_{a}^{\sigma}\left(\mathbf{p};\left[\frac{1}{2},0\right],\frac{1}{2}^-\right) \bar{u}^{b}_{\sigma}\left(\mathbf{p};\left[\frac{1}{2},0\right],\frac{1}{2}^-\right).
	\label{eq:irten_so3_ex_120_012_proj_tensor}
\end{align}
By recalling eqs.~\eqref{eq:spin_wave_func_CGCs}, \eqref{eq:self_conj_rep_spin_wave_func} and \eqref{eq:spin_wave_func_parity_eigenstate}, one can obtain
\begin{equation}
	P_a^{~~b}\left(\mathbf{p};\left[\frac{1}{2},0\right],\frac{1}{2}^+\right) ~=~ \left(\frac{\mathbbm{1}+v_\mu\gamma^\mu}{2}\right)_a^{~~b},\quad
	P_a^{~~b}\left(\mathbf{p};\left[\frac{1}{2},0\right],\frac{1}{2}^-\right) ~=~ \left(\frac{\mathbbm{1}-v_\mu\gamma^\mu}{2}\right)_a^{~~b},
	\label{eq:irten_so3_120_012_proj_func}
\end{equation}
where $v_\mu$ denotes the four-velocity of the particle.
The two spin projection tensors given in  eq.~\eqref{eq:irten_so3_120_012_proj_func} imply Dirac equation.

Considering another kind of spin projection tensors with $[\alpha_1]=[\mu^{s_1}]$, $[\alpha_2]=[\mu^{s_2}]$ and $[\beta]=[\mu^{s}]$, and employing eq.~\eqref{eq:spin_proj_ten_three_chi}, we can derive
\begin{equation}
	P^{\mu^{s_1}\mu^{s_2}}_{\mu^{s}}(\mathbf{p};\chi,s;\chi_{_1},s_1;\chi_{_2},s_2) ~=~ \left(C_s^{s_1s_2}\right)_\sigma^{\sigma_1\sigma_2}\,u^{\sigma}_{\mu^{s}}(\mathbf{p};\chi,s)\, \bar{u}^{\mu^{s_1}}_{\sigma_1}\left(\mathbf{p};\chi_{_1}^*,s_1\right)\,\bar{u}^{\mu^{s_2}}_{\sigma_2}\left(\mathbf{p};\chi_{_2}^*,s_2\right),
	\label{eq:irten_so3_boson_coupling}
\end{equation}
where $\chi=\left[\mu^s\right]$, $\chi_{_1}=\left[\mu^{s_1}\right]$ and $\chi_{_2}=\left[\mu^{s_2}\right]$ are fixed. For convenience, one can rewrite the spin projection tensor as $	P^{\mu^{s_1}\mu^{s_2}}_{\mu^{s}}(\mathbf{p};s,s_1,s_2)$.
This spin projection tensor is corresponding to Lorentz covariant coupling structure among three bosons.

For example, considering the case of $s=s_1=s_2=1$, according to eq.~\eqref{eq:irten_so3_boson_coupling}, one obtains
\begin{equation}
	P^{\nu\rho}_{\mu}(\mathbf{p};1,1,1) ~=~ \frac{i}{\sqrt{2}}~ g_{\mu\mu'} ~v_{\nu'}~ \epsilon^{\mu'\nu'\nu\rho}.
\end{equation}

The derivation of any other IRT of the little group SO(3) follows a similar procedure. These IRTs serve as the building blocks for constructing Lorentz covariant partial wave amplitudes.


\section{Relation between the different forms of spin wave functions}
\label{appd_relation_spin_wave_func}

The conventional form of relativistic spin wave functions  only contains two kinds of Lorentz covariant indices --- Dirac spinor indices $a,b,c,\cdots$ and Lorentz four-vector indices $\mu,\nu,\rho,\cdots$. 
These spin wave functions for spin-0 (time-like four momentum), spin-1 (polarization vector) and spin-$\frac{1}{2}$ (Dirac spinor) are as follows,
\begin{align}
	p_{\mu}\left(\mathbf{p}\right) ~&=~ D_\mu^{~~\mu'}\left(h_\mathbf{p}\right)~k_{\mu'}\left(\mathbf{k}\right),\qquad
	\epsilon^{\sigma}_{\mu}\left(\mathbf{p}\right) ~=~ D_\mu^{~~\mu'}\left(h_\mathbf{p}\right)~\epsilon^{\sigma}_{\mu'}\left(\mathbf{k}\right),\notag\\
	u^{\sigma}_{a}\left(\mathbf{p}\right) ~&=~ D_a^{~~a'}\left(h_\mathbf{p}\right)~u^{\sigma}_{a'}\left(\mathbf{k}\right),\qquad
	v^{\sigma}_{a}\left(\mathbf{p}\right) ~=~ D_a^{~~a'}\left(h_\mathbf{p}\right)~v^{\sigma}_{a'}\left(\mathbf{k}\right),
	\label{eq:appd_commonly_spin_wave_func_motion}
\end{align}
where $D_a^{~~a'}\left(h_\mathbf{p}\right)$ and $D_\mu^{~~\mu'}\left(h_\mathbf{p}\right)$ are pure boost transformations (eq.~\eqref{eq:appd_pure_boost_any_irrep}); 
$u^{\sigma}_{a}\left(\mathbf{k}\right)$, $v^{\sigma}_{a}\left(\mathbf{k}\right)$ and $\epsilon^{\sigma}_{\mu}\left(\mathbf{k}\right)$ are the spin wave functions with standard momentum $k_\mu=[\text{mass}]\cdot\hat{k}_\mu$ as follows,
\begin{align}
	&u^{\sigma}_{a}\left(\mathbf{k}\right) = \begin{pmatrix}
		~\frac{1}{\sqrt{2}}~ & 0 \\
		0 & ~\frac{1}{\sqrt{2}}~ \\
		\frac{1}{\sqrt{2}} & 0 \\
		0 & \frac{1}{\sqrt{2}}
	\end{pmatrix}_a^{~~\sigma},\qquad\quad~\,
	v^{\sigma}_{a}\left(\mathbf{k}\right) = \begin{pmatrix}
		\frac{1}{\sqrt{2}} & 0 \\
		0 & \frac{1}{\sqrt{2}} \\
		-\frac{1}{\sqrt{2}} & 0 \\
		0 & -\frac{1}{\sqrt{2}}
	\end{pmatrix}_a^{~~\sigma},\notag\\
	&\epsilon^{\sigma}_{\mu}\left(\mathbf{k}\right) = \begin{pmatrix}
		0 & 0 & 0\\
		~\frac{1}{\sqrt{2}}~ & 0 & -\frac{1}{\sqrt{2}} \\
		\frac{i}{\sqrt{2}} & 0 & \frac{i}{\sqrt{2}}  \\
		0 & -1 & 0 
	\end{pmatrix}_\mu^{~~\sigma},\qquad
	\hat{k}_\mu\left(\mathbf{k}\right) = 
	\begin{pmatrix}
		1 \\
		0 \\
		0 \\
		0 
	\end{pmatrix}_\mu.
	\label{eq:appd_commonly_spin_wave_func}
\end{align}

According to table~\ref{tab:irreps_Lip}, Dirac spinor wave function carries the self-conjugate REREP $\left[\frac{1}{2},0\right]$, which can only describe a particle with spin-$\frac{1}{2}$ (eq.~\eqref{eq:select_rule_spin_wave_func}); 
while Lorentz four-vector spin wave function carries the self-conjugate IRREP $\left(\frac{1}{2},\frac{1}{2}\right)$, which can describe a particle with spin-0 or spin-1. 
Therefore, to describe a particle with higher spin in the conventional form, one needs to consider the direct product of these two representations, i.e., consider spin wave functions with many Dirac spinor and Lorentz four-vector indices.

In the case of a particle with spin-$\frac{3}{2}$, there are two different ways to describe the respective spin wave function in the following form,
\begin{align}
	[a]\otimes[\mu] =~~~& \left[\left(\frac{1}{2},0\right)\oplus\left(0,\frac{1}{2}\right)\right]\otimes\left(\frac{1}{2},\frac{1}{2}\right) \notag\\
	=~~~& \left(\frac{1}{2},0\right)\oplus\left(0,\frac{1}{2}\right)\oplus\left(1,\frac{1}{2}\right)\oplus\left(\frac{1}{2},1\right),\notag\\
	[a]\otimes[a]\otimes[a] =~~~& \left[\left(\frac{1}{2},0\right)\oplus\left(0,\frac{1}{2}\right)\right]\otimes\left[\left(\frac{1}{2},0\right)\oplus\left(0,\frac{1}{2}\right)\right]\otimes\left[\left(\frac{1}{2},0\right)\oplus\left(0,\frac{1}{2}\right)\right]\notag\\
	=~~~&\left(1,\frac{1}{2}\right)\oplus\left(\frac{1}{2},1\right)\oplus \left(1,\frac{1}{2}\right)\oplus\left(\frac{1}{2},1\right)\oplus \left(1,\frac{1}{2}\right)\oplus\left(\frac{1}{2},1\right)\notag\\ \oplus\,&\left(\frac{3}{2},0\right)\oplus\left(0,\frac{3}{2}\right)\oplus\cdots,
	\label{eq:appd_direct_prod_decomposetion_two_ways}
\end{align}
where the self-conjugate REREP $\left[1,\frac{1}{2}\right]$ belonging to $[a]\otimes[\mu]$ (or $[a]\otimes[a]\otimes[a]$) can describe a particle with spin-$\frac{1}{2}$ or spin-$\frac{3}{2}$; 
the self-conjugate REREP $\left[\frac{3}{2},0\right]$ belonging to $[a]\otimes[a]\otimes[a]$ can describe a particle with spin-$\frac{3}{2}$. 
If one chooses the direct product representation $[a]\otimes[\mu]$ to express the spin wave function, one obtains the following relativistic equations,
\begin{equation}
	\left(\gamma^\mu\right)_a^{~~b}\,u^\sigma_{b\mu}\left(\mathbf{p}\right)=0,\qquad
	\left[p^\nu\,\gamma_\nu-m\,\mathbbm{1}  \right]_a^{~~b}\,u^\sigma_{b\mu}\left(\mathbf{p}\right)=0.
	\label{eq:appd_Rarita_Shcwinger}
\end{equation}
The left hand side of eq.~\eqref{eq:appd_Rarita_Shcwinger} removes 4 components belonging to the REREP $\left[\frac{1}{2},0\right]$ of $[a]\otimes[\mu]$, and the right hand side of eq.~\eqref{eq:appd_Rarita_Shcwinger} removes the other 8 components belonging to the REREP $\left[1,\frac{1}{2}\right]$ of $[a]\otimes[\mu]$.
Then the remaining 4 components of $u^\sigma_{a\mu}\left(\mathbf{p}\right)$ correspond to the 4 polarization components of a particle with spin-$\frac{3}{2}$.
Eq.~\eqref{eq:appd_Rarita_Shcwinger} also implies $p^\mu\,u^\sigma_{a\mu}\left(\mathbf{p}\right)=0$.

The form adopted by Rarita and Schwinger~\cite{Rarita:1941mf} corresponds to the choice described above.
In the form adopted by covariant $L$-$S$ scheme (eq.~\eqref{eq:self_conj_rep_spin_wave_func}), the spin wave function for a particle with spin-$\frac{3}{2}$ are as follows (with conventions in table~\ref{tab:symbol_conventions}),
\begin{equation}
	u^{\sigma}_{a^3}\left(\mathbf{p};\frac{3}{2}^\pm\right) ~=~ \frac{1}{\sqrt{2}} \left[u^{\sigma}_{a^3}\left(\mathbf{p};\left(1,\frac{1}{2}\right),\frac{3}{2}\right) \pm u^{\sigma}_{a^3}\left(\mathbf{p};\left(\frac{1}{2},1\right),\frac{3}{2}\right)\right],
	\label{eq:appd_spin_wave_function_32_a3}
\end{equation}
which can be transformed into Rarita-Schwinger spin wave functions as follows,
\begin{equation}
	u^\sigma_{a\mu}\left(\mathbf{p}\right) ~=~ \left[\left(T_L\right)_{a\mu}^{a^3}+\left(T_R\right)_{a\mu}^{a^3}\right]u^{\sigma}_{a^3}\left(\mathbf{p};\frac{3}{2}^\pm\right)~\equiv~ T_{a\mu}^{a^3}~u^{\sigma}_{a^3}\left(\mathbf{p};\frac{3}{2}^\pm\right),
\end{equation}
where $\left(T_L\right)_{a\mu}^{a^3}$ and $\left(T_R\right)_{a\mu}^{a^3}$ are order-3 IRTs as shown in eq.~\eqref{eq:irten_12+12x1212}.
This relation can be extended to higher-spin cases as follows,
\begin{align}
	&\text{for spin-$\frac{5}{2}$ : }~~~\,u^\sigma_{a\mu\nu}\left(\mathbf{p}\right)\, ~=~ T^{a^3}_{a\mu}~T^{a^5}_{a^3\nu}~ u^{\sigma}_{a^5}\left(\mathbf{p}\right),\notag\\
	&\text{for spin-$\frac{7}{2}$ : }~~~u^\sigma_{a\mu\nu\rho}\left(\mathbf{p}\right) ~=~ T^{a^3}_{a\mu}~T^{a^5}_{a^3\nu}~T^{a^7}_{a^5\rho}~ u^{\sigma}_{a^7}\left(\mathbf{p}\right),
\end{align}
and so on. For spin wave function of a particle with integer-spin, the form adopted by covariant $L$-$S$ scheme (eq.~\eqref{eq:spin_wave_func_irrep}) is related with the conventional form as follows,
\begin{align}
	&\text{for spin-$1$ : }~~~~\epsilon^\sigma_{\mu}\left(\mathbf{p}\right)~\, ~=~ u^{\sigma}_{\mu}\left(\mathbf{p};[\mu],1\right),\notag\\
	&\text{for spin-$2$ : }~~~\,\epsilon^\sigma_{\mu\nu}\left(\mathbf{p}\right)\,\hspace{0.3mm} ~=~ T^{\mu^2}_{\mu\nu}~u^{\sigma}_{\mu^2}\left(\mathbf{p};\left[\mu^2\right],2\right),\notag\\
	&\text{for spin-$3$ : }~~~\epsilon^\sigma_{\mu\nu\rho}\left(\mathbf{p}\right) ~=~ T^{\mu^2}_{\mu\nu}~T^{\mu^3}_{\mu^2\rho}~u^{\sigma}_{\mu^3}\left(\mathbf{p};\left[\mu^3\right],3\right),
\end{align}
and so on.

If the direct product representation $[a]\otimes[a]\otimes[a]$ is chosen to express the spin wave function, the following relativistic equations are obtained for a particle with spin-$\frac{3}{2}$,
\begin{equation}
	\left[p^\mu\,\gamma^k_\mu-m\,\mathbbm{1}  \right]_{a_k}^{~~b_k}\,u^\sigma_{b_1b_2b_3}\left(\mathbf{p}\right)=0~~(k=1,2,3),
	\label{eq:appd_Wigner}
\end{equation}
where $\gamma^k_\mu~(k=1,2,3)$ are Dirac-$\gamma$ matrices of $k$-th index $b_k$ and three Dirac spinor indices of $u_{b_1b_2b_3}\left(\mathbf{p}\right)$ are symmetric.
Eq.~\eqref{eq:appd_Wigner} reduces the 64 components belonging to the  representation $[a]\otimes[a]\otimes[a]$ to 8 components, while the symmetrization of the three Dirac indices $b_1,b_2,b_3$ further reduces to 4 components which correspond to the 4 polarization components of a particle with spin-$\frac{3}{2}$.
This choice corresponds to the form adopted by Bargmann and Wigner \cite{Bargmann:1948ck}. 

In the form adopted by covariant $L$-$S$ scheme (eq.~\eqref{eq:self_conj_rep_spin_wave_func}), the Bargmann-Wigner spin wave function for a particle with spin-$\frac{3}{2}$ are given as follows,
\begin{equation}
	u^\sigma_{a_1a_2a_3}\left(\mathbf{p}\right) ~=~ \frac{1}{2\sqrt{2}} \sum_{[\alpha]\in[a]\otimes[a]\otimes[a]} T_{a_1a_2a_3}^{\alpha} u^{\sigma}_{\alpha}\left(\mathbf{p};[\alpha],\frac{3}{2}\right),
	\label{eq:appd_relation_W_irten}
\end{equation}
where the summation of $[\alpha]$ span all possible IRREP belonging to the direct product representation $[a]\otimes[a]\otimes[a]$, which has 8 possibilities as shown in eq.~\eqref{eq:appd_direct_prod_decomposetion_two_ways}, and $T_{a_1a_2a_3}^{\alpha}$ is an order-4 IRT of $L_p$ which can be constructed by order-3 IRTs (eq.~\eqref{eq:irten_order_3_CGCs}).
The spin wave function in eq.~\eqref{eq:appd_relation_W_irten} is for particle with positive parity. Similar to eq.~\eqref{eq:appd_spin_wave_function_32_a3}, the spin wave function with negative parity can be obtained by changing the relative sign between the spin wave functions on the right hand side, which are conjugate representations of each other.

To summarize, an IRREP $[\alpha]=\left(s_L,s_R\right)$ can be utilized to represent a particle with spin-$s$ provided that it satisfies the triangular relation (eq.\eqref{eq:select_rule_spin_wave_func}). Consequently, there exist numerous ways to describe a particle with a definite spin. These different forms of spin wave functions can be transformed into each other via the use of IRTs of $L_p$.


\section{Explicit form of the weight function \boldmath{$W(s_1,s_2,s_3,L,S)$}}
\label{appd:weight_func}

For processes involving one massless particle, the weight function is given by:
$$W(s_1,s_2,s_3,L,S)=F_{S}(s_1,s_2,s_3,S).$$
For processes involving two or three massless particles, the weight function is given by:
$$
W(s_1,s_2,s_3,L,S)=F_{S}(s_1,s_2,s_3,S)+F_{L}(s_1,s_2,s_3,L,S)+F_{\sigma}(s_1,s_2,s_3,L,S),
$$
where the individual weight functions are defined as follows:
\begin{align}
	F_{S}(s_1,s_2,s_3,S)~&=~-(s_2+s_3+1)\left|S-s_1\right|+S,\notag\\
	F_{L}(s_1,s_2,s_3,L,S)~&=~-2(s_2+s_3+1)^2\left|L-|S-s_1|-\frac12\right|,\notag\\
	F_{\sigma}(s_1,s_2,s_3,L,S)~&=~\left\{\begin{array}{cl}
		-2(s_2+s_3+1)^2(s_1+s_2+s_3) ~&~\text{for~} \left(C_{s_1}^{LS}\right)_{s_2\pm s_3}^{0\,s_2\pm s_3}=0\\
		&\\
		0 & ~\text{for others}
	\end{array}\right..\notag
\end{align}


\bibliography{refs}
\end{document}